\newif\iftechrep
\techreptrue
\newif\ifnottechrep
\iftechrep\nottechrepfalse\else\nottechreptrue\fi
\newif\ifusenix
\usenixfalse

\iftechrep
  \documentclass[sigplan,10pt,authorversion]{acmart}
\else 
  \ifusenix
	\documentclass[letterpaper,twocolumn,10pt]{article}
	\usepackage{usenix2019_v3,epsfig,endnotes}
	\usepackage{tikz}
	\usepackage{amsmath}
	\usepackage{filecontents}
  \else
	\documentclass[sigplan,10pt]{acmart}
  \fi
\fi

\usepackage{tikz}
\usetikzlibrary{calc}
\usetikzlibrary{chains}
\usepackage{filecontents}
\usepackage{ifthen}

\usepackage{booktabs}
\usepackage{multirow}
\usepackage{xspace}
\usepackage{xcolor}
\usepackage{amsmath,latexsym,amssymb,amsthm,amsfonts}
\usepackage{pifont}
\usepackage{booktabs}

\usepackage{subfigure}

\PassOptionsToPackage{hyphens}{url}\usepackage{hyperref}

\hypersetup{
	colorlinks=true,
	linkcolor=blue,
	citecolor=red,
	urlcolor=magenta,
}

\newcommand{\cref}[1]{{\S\ref{#1}}}

\newboolean{showcomments}
\setboolean{showcomments}{true}

\ifthenelse{\boolean{showcomments}}{%
  \newcommand{\mynote}[2]{%
    {\small\sffamily{\color{red}#1: \color{blue} #2}}}
  \newcommand{\todo}[1]{%
    {\small\textbf{\color{red}#1}}}
}{%
  \newcommand{\mynote}[2]{}
  \newcommand{\todo}[1]{}
}
          
\newcommand{\gauthier}[1]{}
\newcommand{\vincent}[1]{}
           
\newcommand{\bftmencius}{BFT-Mencius\xspace}
\newcommand{\bftsmart}{BFT-SMaRt\xspace}
\newcommand{\epaxos}{EPaxos\xspace}
\newcommand{\ezbft}{ezBFT\xspace}
\newcommand{\hotstuff}{HotStuff\xspace}
\newcommand{\paxos}{Paxos\xspace}
\newcommand{\prsm}{{\sc Dispel}\xspace}

\newcommand{\zyzzyva}{Zyzzyva\xspace}

\newcommand{\smr}{SMR\xspace}

\newcommand{\code}[1]{\texttt{#1}}

\usepackage{algorithmicx}
\usepackage{algpseudocode}
\usepackage{algorithm}
\usepackage{macros}
\usepackage{ioa_code}

\newenvironment{smallenum}{
\begin{enumerate}
    \setlength{\topsep}{-3pt} 
    \setlength{\partopsep}{-3pt}
  \setlength{\itemsep}{0pt}
  \setlength{\parskip}{-1pt}
  \setlength{\parsep}{-6pt}
}{\end{enumerate}}

\begin{document}

\iftechrep
  \setcopyright{none}
  \makeatletter
  \renewcommand\@formatdoi[1]{\ignorespaces}
  \makeatother
  \title[\prsm]{{\prsm}: Byzantine SMR with Distributed Pipelining} 
  \author{Gauthier Voron} 
  \orcid{}
  \affiliation{University of Sydney}
  \email{gauthier.voron@sydney.edu.au}
  \author{Vincent Gramoli} 
  \orcid{0000-0001-5632-8572}
  \affiliation{University of Sydney}
  \email{vincent.gramoli@sydney.edu.au}
    \settopmatter{printacmref=false}
    \setcopyright{none}
    \renewcommand\footnotetextcopyrightpermission[1]{}
    \pagestyle{plain}
\fi


\begin{abstract}
Byzantine State Machine Replication (\smr) is a long studied topic that received increasing attention recently with the advent of blockchains as companies are trying to scale them to hundreds of nodes. 
Byzantine \smr{}s try to increase throughput by either reducing the latency of consensus instances that they run sequentially or by reducing the number of replicas  that send messages to others in order to reduce the network usage. 
Unfortunately, the former approach makes use of resources in burst whereas the latter requires CPU-intensive authentication mechanisms.

In this paper, we propose a new Byzantine \smr called \prsm (\emph{\underline{Dis}tributed \underline{P}ip\underline{el}ine}) that allows any node to distributively start new consensus instances whenever they detect sufficient resources locally. 
We evaluate the performance of \prsm within a single datacenter and across up to 380 machines over 3 continents by comparing it against four other \smr{}s. 
On 128 nodes, \prsm speeds up 
HotStuff, the Byzantine fault tolerant SMR being integrated within Facebook's blockchain, by more than 12 times.
In addition, we also test \prsm under isolated and correlated failures and show that
the \prsm distributed design is more robust than HotStuff.
Finally, we evaluate \prsm in a cryptocurrency application with Bitcoin transactions and show that this \smr{} is not the bottleneck.
\end{abstract}

\iftechrep
\pagestyle{plain}
\settopmatter{printfolios=true}
\maketitle
\fi

\section{Introduction}
\label{section:introduction}

State machine replication (\smr) makes use of consensus for 
totally ordering a set of commands (or proposals) that are executed in the same 
order by all replicas. Consensus protocols are generally 
network bound as they often rely on some broadcast 
patterns to minimize the time it takes to reach agreement
on the next command. 
Byzantine fault tolerant (BFT) \smr{}s have regained in popularity with the introduction of blockchain technology:
Facebook even aims at deploying a variant of the \hotstuff \smr~\cite{YMR19} on at least 100 replicas over a large network~\cite[\S{}5]{Fac19}.
With the 1.6 billion daily active users on Facebook\footnote{\url{https://newsroom.fb.com/company-info/}.}, comes the question of the amount of payload 
needed to be treated by such a blockchain system once it will be in production.
Unfortunately, the performance of \smr{}s generally drops significantly before reaching a hundred of nodes.

One of the reasons of this limitation is commonly believed to be
the all-to-all communication pattern between replicas~\cite{Vuc15,CJB18,YMR19}. In fact, 
$n$ replicas sending messages to all other replicas necessarily lead to $\Theta(n^2)$ 
messages~\cite{CL02,CWA09,BSA14}.
Given that the network bandwidth is a limited resource, it could seem that 
this quadratic complexity becomes unaffordable on large networks, like the Internet.
For these reasons, various protocols~\cite{AMG05,KAD07,AGK14,YMR19,BBC19} 
replaced this all-to-all message exchanges by one-to-all exchanges where they could. 
The problem is that the 
evaluation of network usage 
is far from being trivial and unexpected causes may impact the observed throughput.

\begin{figure}[t]
\begin{center}
\includegraphics[scale=0.48,clip=true,viewport= 0 0 500 225]{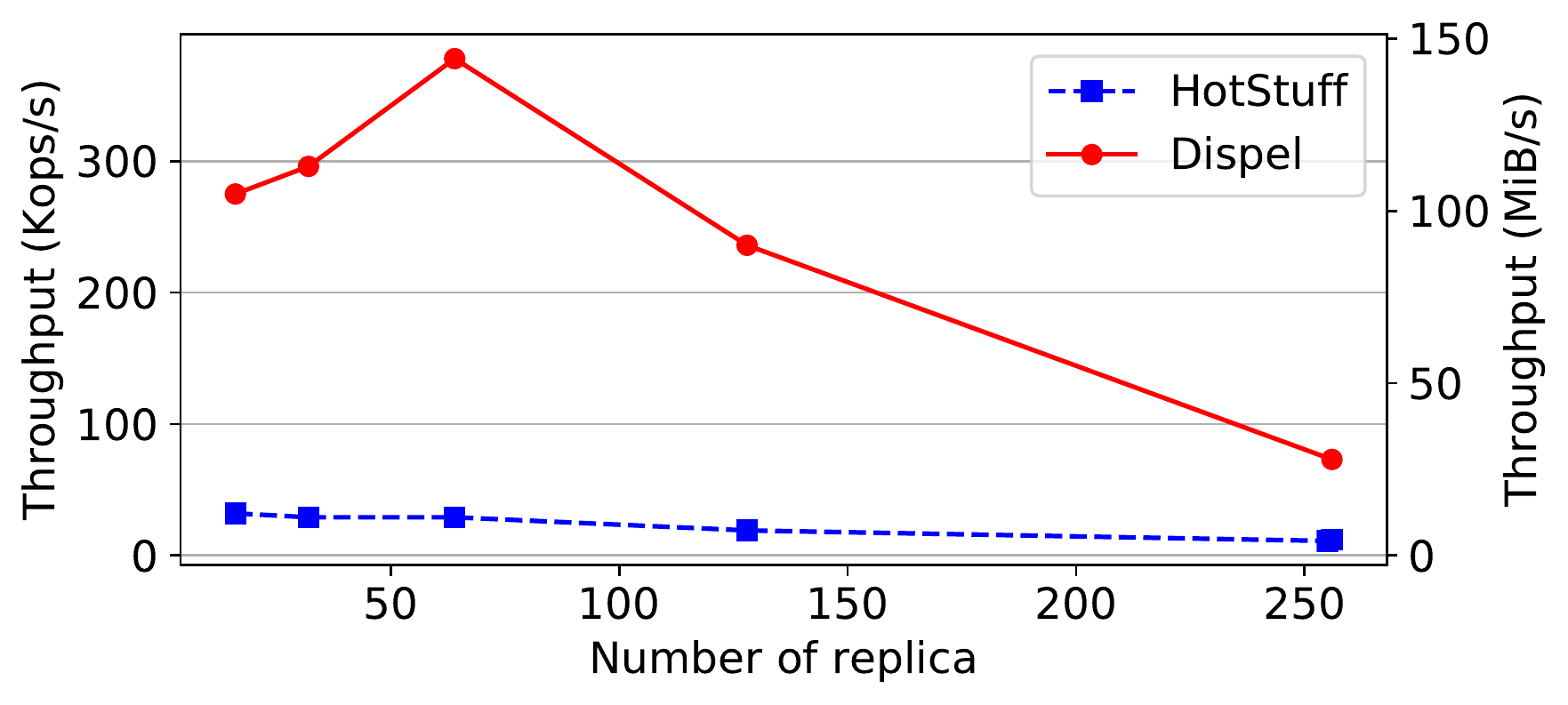}
\caption{The goodput of \smr{}s (e.g., \hotstuff) 
is low due to its unbalanced bandwidth consumption
whereas \prsm exploits the bandwidth  of more links to offer, at 128 nodes, a 12-fold speedup.
\label{fig:scale-throughput-intro}}
\end{center}
\ifnottechrep\vspace{-0.5em}\fi
\end{figure}	

In this paper, we revisit this idea by offering a new \smr{}, called \prsm (\emph{\underline{Dis}tributed \underline{P}ip\underline{el}ine}) that balances its quadratic number of messages onto as many routes between distributed replicas to offer high throughput with reasonably low latency.
\prsm follows from a year-long investigation of the application-level throughput in \smr{}s---commonly referred to as goodput that accounts for payload amount.
This extensive evaluation of network performance
allowed us to identify 
important causes of performance limitations, like \emph{head-of-line (HOL)} blocking~\cite{SK06} 
that delays subsequent packet delivery due to one TCP packet loss.
To illustrate the benefit of \prsm{} consider Figure~\ref{fig:scale-throughput-intro} that compares its throughput to \hotstuff, the latest \smr that we are aware of. Although \hotstuff outperforms preceding \smr{}s thanks to its linear message complexity, it suffers from the same leader-based message pattern as its predecessors (cf. 
\cref{section:evaluation} for the detailed setting of this figure and more results).

The key innovation of \prsm is its distributed pipeline,
%
%
a technique
adapted from the centralized pipelining of micro-architectures to 
the context of \smr to leverage unused resources.
Both pipelining techniques consist of executing multiple \emph{epochs} (or concurrent executions) 
in order to provide higher throughput than what could be obtained
with a single epoch.
As opposed to the classic pipelining that aims at 
maximizing the usage of central resources, the distributed pipelining 
maximizes the resource usage of distributed replicas by allowing them
to decide locally when to spawn new consensus epochs.
In particular, in \prsm each replica that detects idle network 
resources at its network interface (NIC) and sufficient available memory 
spawns a new consensus
instance in a dedicated pipeline epoch. This distributed detection is 
important to leverage links of heterogeneous capacity like inter-datacenters 
vs. intra-datacenters communication links (cf.~\cref{ssec:hol}).
We draw three conclusions out of this work:
\begin{smallenum}


\item {\bf HOL blocking limits \smr{}s at high speed.} 
\prsm allows us to increase the performance of the \smr to a new level, where 
we identified head-of-line (HOL) blocking as the bottleneck (instead of the network interface (NIC) bandwidth). 
This phenomenon is detailed in \cref{ssec:hol} where the throughput increases at small scale 
due to the multiplication of communication routes between replicas 
and
where 
performance increases proportionally to the number of TCP connections. 

\item {\bf Distributed pipelining increases robustness.}
Distributing the pipelining allows any replica to start a new consensus instance by proposing requests, hence allowing the \smr{} to progress despite the failure of a single replica. 
This is in contrast with centralized pipelining: 
the leader failure can impact the \smr{} performance until a correct leader is selected.
This is detailed in~\cref{ssec:robust} where a single failure has no noticeable impact on \prsm
and where correlated failures do not prevent \prsm from treating 150,000 requests per second.

\item {\bf CPU-bound planet-scale BFT cryptocurrency.} Our experiments of \prsm within a cryptocurrency application where miners verify all transactions (as in classic blockchains~\cite{Nak08,Woo15}) show that the performance is limited locally by the cryptographic cost at each node and no longer globally by the network cost of BFT consensus. 
A direct consequence is that the performance does not drop when the system size increases, even when deployed on up to 380 nodes distributed across 3 continents (\cref{sec:blockchain}).
\end{smallenum}

In the remainder of the paper, we start by presenting the background that motivates our work (\cref{section:background}). 
Then we present the design of \prsm to pipeline the consensus executions (\cref{section:design}) before we detail the implementation choices (\cref{section:implementation}).
We then evaluate \prsm within a datacenter, on geo-distributed systems across 3 continents, against other \smr{}s (\cref{section:evaluation}), under correlated failures, and when running a cryptocurrency application (\cref{sec:blockchain}). 
Finally, we discuss the related work (\cref{section:rw}) and  conclude (\cref{section:conclusion}).


\section{Background}
\label{section:background}

BFT state machine replication (SMR) relies on executing repeatedly a BFT consensus protocol  designed for subsecond latency. Typical implementations are sequential~\cite{CL02,KAD07,BSA14,AGK14}; they execute one consensus instance at a time. 
In order to increase throughput, they batch commands (e.g.,~\cite{CL02,BSA14}), hence proposing multiple commands to the same consensus instance.
More precisely, when the system starts, $n$ replicas (or \emph{nodes}) propose a batch (or \emph{proposal}), possibly empty. They execute a BFT consensus instance until they decide on a batch. 
When a batch is decided, the replicas execute all the commands of this batch and then start a new 
consensus instance.
Executing consensus instances sequentially is instrumental in identifying and discarding conflicting commands in the batch they process.
This sequential design, yet simple, is efficient enough when the consensus latency is low.

BFT consensus instances totally order proposals at consecutive indices.
As it is impossible to implement consensus in an asynchronous failure-prone environment~\cite{FLP85}, 
various systems assume a \emph{partially synchronous} environment, in that every message gets delivered in an unknown but bounded amount of time~\cite{DLS88}. In this environment, one can solve the BFT consensus if the number $f$ of \emph{Byzantine} (or malicious) nodes is lower than $\frac{n}{3}$~\cite{PSL80}, meaning that $n-f$ nodes remain \emph{correct}.
Most of the practical BFT \smr{}s implemented today 
rely on a leader-based pattern 
where a single node exchanges messages with Byzantine quorums of $2f+1$ nodes. 
Recently, BFT \smr gained attention for its ability to totally order blocks of transactions in a blockchain~\cite{Nak08} in which the challenge becomes to perform efficiently on larger networks.

\subsection{The network bottleneck of the first phase}\label{ssec:leader-bottleneck}
%
The leader-based pattern typically starts with a message exchange \emph{phase} where a specific node, called the \emph{leader}, aims at proposing a command to the rest of the nodes. 
If the leader is faulty, the system may choose another leader.
%
As it is impossible to distinguish a slow network from a mute leader in a partially synchronous environment, such changes affect performance~\cite{SRM12,MAK13,AMQ13,GPB16}.
If the leader is correct and the network is timely, then the command is decided by all nodes.
But in this case, the leader may have to send its proposal to many, making its NIC the bottleneck~\cite{JRS11}.
One may think of batching
proposals within the same consensus instance~\cite{FH06}, so that multiple proposals are piggybacked in the same messages and can be decided in a row. This increases the information conveyed to all nodes by the leader, hence adding to the congestion.
%

\subsection{The CPU-intensive subsequent phases}
\label{ssec:cpu-instensive}
To mitigate the leader bottleneck once the proposals are conveyed to all nodes, the nodes typically hash the content of the proposal into some digest~\cite{CL02,BSA14,YMR19}
and exchange the resulting digests to refer to specific proposals. On the one hand, this reduces considerably the network utilization in 
the phases following the prepare phase.
On the other hand, the subsequent phases convey more frequent but smaller messages than in the first phase whose treatment consumes CPU.
The hashing function
necessary to encode these messages 
is also CPU intensive. As the communication is partially synchronous it is likely 
that the hashing function execution overlaps at many nodes with the reception of these message digests, hence further increasing the CPU usage.

When requests must be cryptographically verified~\cite{BSA14}, as in 
cryptocurrency applications~\cite{Nak08,Woo15}, 
or when the phases require message authenticators~\cite{CL02}, the system can become CPU 
bound.
To put things in perspective, an AWS EC2 \textit{c5.xlarge}
instance has an upload bandwidth of 600 MiB/s but a CPU of this same instance
can only hash 425 MiB/s with SHA256 and verify up to 5\,MiB/s for 400 byte
transactions with the fastest ECDSA curve provided by OpenSSL as we detail in~\cref{sec:blockchain}.
BFT \smr{}s typically alternate between the network-intensive phase and 
these CPU-intensive phases.

\subsection{Bypassing network and CPU bottlenecks}
\label{ssec:dbft}

To bypass the leader completely, several deterministic leaderless Byzantine consensus were proposed.
Lamport suggests a virtual leader election~\cite{Lam10,Lam11} to transform a leader-based Byzantine consensus algorithm into a leaderless one, however, no virtual leader election algorithm is given.
Borran and Schiper~\cite{BS10} proposed a Byzantine leaderless consensus whose communication complexity is exponential.

Recently, Crain et al.~\cite{CGLR18} proposed a leaderless deterministic Byzantine consensus algorithm.
The algorithm called Democratic BFT (DBFT), is 
%
%
depicted in Algorithm~\ref{algo:dbft} 
and builds upon a reduction of the problem of multivalue consensus to the problem of binary consensus~\cite{BKR94} and a resilient-optimal and time-optimal deterministic binary consensus algorithm that was formally verified using model checking~\cite{TG19}: Each replica of id $0\leq k<n-1$ reliably broadcasts its input value such that all correct replicas deliver the same value~\cite{B87} into the $k^{th}$ coordinate of a local array. Later, DBFT spawns $n$ binary instances whose $k^{th}$ takes input value 1 if the $k^{th}$ coordinate of the array was reliably delivered, or 0 if it was not yet delivered. The decisions of the $n$ binary instances form a bitmask that is applied to the array of values to extract the decidable values. DBFT outputs the first of these values to solve the classic Byzantine consensus problem, the Red Belly Blockchain extends DBFT~\cite{CNG18} to output all these values hence committing more transactions, unfortunately, it runs consensus instances sequentially. 

\begin{algorithm}[t]
	\caption{DBFT, Consensus for Blockchains~\cite{CGLR18}\label{algo:dbft}
  }
  
  \renewcommand{\Comment}[1]{}
	{\footnotesize

	\begin{algorithmic}[1]
	
		\Part{$\lit{consensus}(\ms{batch})$} \label{line:cons-start}  {\color{gray}{\textit{// similar to \cite{BKR94}}}}
			\State $\lit{reliable-broadcast}(\ms{batch})$\label{line:rbcast}  {\color{gray}{\textit{// e.g., \cite{B87}}}}
			\State $\ms{array}[1..n] \gets$ reliably-delivered $n-t$ batches\label{line:threshold} {\color{gray}{\textit{// wait for $n-t$ values}}}
			\For{$k=1..n$}
				\If{$\ms{array}[k]$ was reliably delivered}
					\State $\ms{bitmask}[k] \gets \lit{fast-deterministic-binary-consensus}_k(1)$\label{line:propose-1}   
				\Else{} $\ms{bitmask}[k] \gets \lit{fast-deterministic-binary-consensus}_k(0)$\label{line:propose-0}  
				\EndIf
			\EndFor
				\WUntil{$\ms{bitmask}$ is full and $\forall \ell,\ms{bitmask}[\ell] = 1 : \ms{array}[\ell] \neq \emptyset$}\label{line:bitmask}\EndWUntil 
			\Return ($\ms{bitmask}  ~\&~ \ms{array}$)  {\color{gray}{\textit{// apply the bitmask to decide a batch subset}}}
			\EndReturn \label{line:cons-end}
		\EndPart

	\end{algorithmic}
	}
\end{algorithm}

\subsection{Limits of sequential consensus instances}
Once the bottleneck effects are mitigated, one can further improve the throughput of \smr{}s by reducing the latency of 
one consensus instance or by overlapping different consensus instances.
First, a long series of results~\cite{FV97,KAD07,SR08,SB12,MBS13,VCB13} reduce 
the latency of the BFT consensus, sometimes by assuming correct clients~\cite{KAD07}, tolerating less Byzantine nodes~\cite{SR08} or using a trusted execution environment~\cite{VCB13,KBC12}. 
The effect of reducing latency on increasing the throughput 
is quite visible~\cite{MBS13}, however, this increase is limited by the sequential execution of these consensus instances. 

\sloppy{Second, some form of centralized pipelining was proposed~\cite{SS13,YMR19}. This technique
inherited from 
networking~\cite{PM95}, 
consists of executing some consensus instance before another terminates. 
This approach is ``centralized'' because it is leader-based: a leader is needed to spawn a new pipeline epoch.
The benefit of centralized pipelining was observed to be limited~\cite{SS13}, again
due to the network bottleneck at the leader (cf.~\cref{ssec:leader-bottleneck}).
\hotstuff~\cite{YMR19} is a recent BFT \smr that piggybacks phases of consecutive consensus instances
and Facebook is building Libra~\cite{BBC19} on top of its variant.\footnote{\url{https://medium.com/ontologynetwork/hotstuff-the-consensus-\ protocol-behind-facebooks-librabft-a5503680b151}.\label{note1}}
Unfortunately, it again relies on a leader. \prsm is the first distributed pipeline \smr{} as described below.}

\section{\prsm Overview}
\label{section:design}


%

This section presents \prsm, standing for \emph{\underline{Dis}tributed \underline{P}ip\underline{el}ine},  a BFT \smr for the partially synchronous model tolerating $f<n/3$ Byzantine replicas.
Unlike previous \smr{}s, each replica in \prsm spawns a new consensus instance based on its available local resources.

\begin{figure}[t]
  \begin{center}
    \begin{tikzpicture}
  \tikzstyle{component}=[draw, thick, rounded corners, minimum height=18pt,
    minimum width=50pt, font=\small\bf];
  \tikzstyle{external}=[component, fill=black!10]
  \tikzstyle{legend}=[font=\normalsize]
  \tikzstyle{order}=[font=\normalsize]
  \tikzstyle{astep}=[inner sep=0, font=\large\bf]
  \tikzstyle{transfer}=[->, very thick]
  \tikzstyle{tx}=[transfer, black, dashed]
  \tikzstyle{hash}=[transfer, red, densely dotted]
  \tikzstyle{proposal}=[transfer, blue!60, line width=3pt]

  \node[component, minimum width=100pt, anchor=north west] (hasher) at (0, 0) {Hasher};
  \node[component, anchor=north west] (pool) at ($(hasher.south west) + (0, -15pt)$) {Pool};
  \node[external, anchor=north west] (client) at ($(pool.south west) + (0, -25pt)$) {Clients};

  \path (pool.north) -| node (tmp) {} (hasher.east);
  \node[component, minimum width=60pt, anchor=north west] (consensus) at ($(tmp) + (-20pt, 0)$) {Consensus};

  \path (client.north) -| node (tmp) {} (consensus.east);
  \node[external, minimum width=80pt, anchor=north east] (server) at (tmp) {Other Replicas};

  \path (hasher.north) -| node (tmp) {} (consensus.east);
  \draw let \p1=($(tmp.center) - (consensus.south east)$),
            \n2={veclen(\x1,\y1)} in
    node[component, minimum height=\n2-\pgflinewidth, anchor=south west] (manager) at ($(consensus.south east) + (20pt, 0)$) {Manager};

  \node (netw) at ($(pool.south west) + (0, -20pt)$) {};
  \path (netw.center) -| node (nete) {} (manager.east);
  \draw[very thick] (netw.center) -- (nete.center);

  \draw[tx] (client) -- node[astep, anchor=west, xshift=4pt, yshift=4pt] (1) {1} (pool);

  \path (pool.north) |- node (tmp) {} (hasher.south);
  \draw[proposal] (pool.north) -- node[astep, anchor=east, xshift=-6pt] (2a) {2A} (tmp.center);
  \path (pool.north) -- node[yshift=-2pt] (tmp0) {} (hasher.south);
  \path (server.west) -- node (tmp1) {} (consensus.west);
  \path (tmp1.center) -- node (tmp2) {} (server.west);
  \path (tmp2.center) |- node (tmp3) {} (server.north);
  \draw[proposal] (pool.north) |- (tmp0.center) -| (tmp3.center);

  \path (server.west) -- node (tmp) {} (consensus.west);
  \path (tmp.center) -- node (tmp) {} (consensus.west);
  \path (server.north) -| node (tmp0) {} (tmp.center);
  \path (hasher.south) -| node (tmp1) {} (tmp.center);
  \draw[proposal] (tmp0.center) -- node[anchor=west, xshift=9pt] (tmp) {} (tmp1.center);
  \path (tmp.center) |- node (tmp) {} (2a.center);
  \node[astep, blue!60] (2b) at (tmp.center) {2B};

  \path (consensus.south) |- node (tmp) {} (server.north);
  \draw[hash, <->] (tmp.center) -- node[anchor=west, xshift=6pt] (tmp) {} (consensus.south);
  \path (tmp.center) |- node (tmp) {} (1.center);
  \node[astep, red] (2c) at (tmp.center) {2C};

  \path (consensus.east) -| node (tmp) {} (manager.west);
  \draw[hash] (consensus.east) -- node[astep, anchor=south, yshift=2pt] (4) {4} (tmp.center);

  \path (hasher.east) -| node (tmp) {} (manager.west);
  \draw[proposal] ([yshift=3pt]hasher.east) -- ([yshift=3pt]tmp.center);
  \draw[hash] ([yshift=-3pt]hasher.east) -- ([yshift=-3pt]tmp.center);
  \draw[hash] ([yshift=-3pt]hasher.east) -| node[astep, anchor=north, xshift=6pt, yshift=-2pt] (3) {3} (consensus.north);

  \path (nete.center) -| node[yshift=3pt] (tmp) {} (manager.south);
  \draw[proposal] (manager.south) -- node[astep, anchor=west, xshift=3pt, yshift=3pt] (5) {5} (tmp.center);

  \node (legendw) at ($(client.south west) + (0, -12pt)$) {};
  \path (manager.east) |- node (legende) {} (legendw.center);

  \node[legend, anchor=west] (txlegend) at ($(legendw.center) + (15pt, 0)$) {transaction};
  \draw[tx] (legendw.center) -- (txlegend);

  \node (tmp) at ($(txlegend.east) + (15pt, 0)$) {};
  \node[legend, anchor=west, color=blue!60] (proposallegend) at ($(tmp.center) + (15pt, 0)$) {batch};
  \draw[proposal] (tmp.center) -- (proposallegend.west);

  \node (tmp) at ($(proposallegend.east) + (15pt, 0)$) {};
  \node[legend, anchor=west, color=red] (hashlegend) at ($(tmp.center) + (15pt, 0)$) {hash};
  \draw[hash] (tmp.center) -- (hashlegend.west);
\end{tikzpicture}
  \end{center}
  \caption{\label{figure:archi2}Architecture of \prsm.
    Transactions are collected in the pool and batched (1).
    Replicas exchange batches (2A,2B), store and associate them with their hash digest in the manager (3).
    Replicas execute consensus protocol over hashes (2C,3) and transmit the decisions to the manager (4).
    The manager transmits the batches associated with the decided hashes to the application (5).}
\end{figure}
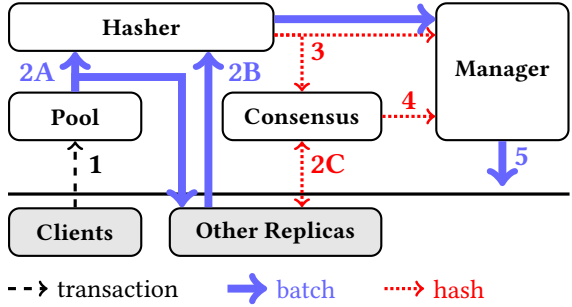

\subsection{Architecture of a pipeline epoch}
\label{ssec:pipeline-stage}
Figure~\ref{figure:archi2} shows the architecture and the main steps of one \emph{pipeline epoch} that runs one consensus instance.
This epoch consists of
creating a batch of commands, called \emph{transactions} in the context of blockchains, exchanging batches with other replicas and selecting an ordered subset of the batches created by the replicas.
A \prsm replica continuously listens for client transactions and stores them in its transaction pool (step 1).
When a replica decides to spawn a new pipeline epoch, it concatenates all the transactions in the pool to create a batch.
The replica then broadcasts the batch as the first step of the reliable broadcast of the DBFT consensus protocol (\cref{ssec:dbft}).
In parallel, a dedicated hashing thread computes the \texttt{sha256} checksum of the batch (step 2A).
%

All \prsm replicas decide independently to spawn a new pipeline epoch.
We describe this process in details in \cref{ssec:phase-details}.
As a consequence, a replica receives batches from the other replicas in parallel to steps 1 and 2A (in step 2B).
When a hashing thread has computed the hash of a batch, it stores the batch and its associated hash in the manager component (step 3).
The hashing thread also transmits the hash to the consensus component.
From this point on, the consensus component sent the hash digest instead of the batch itself when communicating with other replicas.
The consensus component exchanges hashes to complete the reliable broadcast and executes the subsequent steps of Algorithm~\ref{algo:dbft} (step 2C). 
When the replicas decide a set of hashes, the consensus component transmit the decision to the manager (step 4).
The manager then fetches the batches associated to the decided hashes and gives these decided batches to the pipeline epoch orderer that we describe in \cref{ssec:pipeline-order} (step 5).

\subsection{Pipeline overview}
\label{ssec:pipeline-spawn}

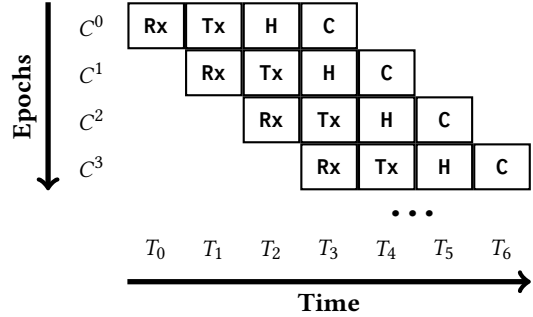
\begin{figure}[t]
  \begin{center}
    \begin{tikzpicture}
  \tikzstyle{phase}=[draw, thick, minimum width=21pt, minimum height=17pt,
    font=\small\bf\ttfamily, inner sep=2pt]
  \tikzstyle{stageid}=[font=\small]
  \tikzstyle{timeid}=[font=\small]
  \tikzstyle{time}=[draw,->,line width=2pt]

  \node[phase] (rx0) {Rx};
  \node[phase, anchor=west] (tx0) at (rx0.east) {Tx};
  \node[phase, anchor=west] (h0) at (tx0.east) {H};
  \node[phase, anchor=west] (c0) at (h0.east) {C};

  \node[phase, anchor=north] (rx1) at (tx0.south) {Rx};
  \node[phase, anchor=west] (tx1) at (rx1.east) {Tx};
  \node[phase, anchor=west] (h1) at (tx1.east) {H};
  \node[phase, anchor=west] (c1) at (h1.east) {C};

  \node[phase, anchor=north] (rx2) at (tx1.south) {Rx};
  \node[phase, anchor=west] (tx2) at (rx2.east) {Tx};
  \node[phase, anchor=west] (h2) at (tx2.east) {H};
  \node[phase, anchor=west] (c2) at (h2.east) {C};

  \node[phase, anchor=north] (rx3) at (tx2.south) {Rx};
  \node[phase, anchor=west] (tx3) at (rx3.east) {Tx};
  \node[phase, anchor=west] (h3) at (tx3.east) {H};
  \node[phase, anchor=west] (c3) at (h3.east) {C};

  \node[anchor=north, font=\huge\bf, minimum height=17pt] (dots) at (tx3.south east) {\ldots};

  \node[stageid, anchor=east] (s0) at ($(rx0.west) + (-5pt, 0)$) {$C^0$};
  \foreach \i in {1,2,3} {
    \path (s0.west) |- node[stageid, anchor=west] (s\i) {$C^\i$} (rx\i.west);
  }

  \path (rx0.south) |- node[timeid, yshift=-5pt] (t0) {$T_0$} (dots.south);
  \foreach \i in {1,2,3} {
    \path (t0.south) -| node[timeid, anchor=south] (t\i) {$T_\i$} (rx\i.south);
  }
  \path (t0.south) -| node[timeid, anchor=south] (t4) {$T_4$} (c1.south);
  \path (t0.south) -| node[timeid, anchor=south] (t5) {$T_5$} (c2.south);
  \path (t0.south) -| node[timeid, anchor=south] (t6) {$T_6$} (c3.south);

  \path (t0.south) -| node[yshift=-5pt] (timestart) {} (rx0.west);
  \path (timestart.center) -| node (timeend) {} (c3.east);
  \draw[time] (timestart.center) -- node[anchor=north, font=\bf] {Time} (timeend.center);

  \node (stagestart) at ($(rx0.north west) + (-30pt, 0)$) {};
  \path (stagestart.center) |- node (stageend) {} (rx3.south);
  \draw[time] (stagestart.center) -- node[anchor=south, rotate=90, font=\bf] {Epochs} (stageend.center);
\end{tikzpicture}
  \end{center}
  \caption{Principle of a consensus pipeline.
    A pipeline epoch consists of four phases: network reception (Rx), network transmission (Tx), CPU intensive hash (H) and latency bound consensus (C).
    As soon as an epoch finishes its first phase (Rx), the replica spawns a new epoch.
    When the replica executes four epochs concurrently, it leverages all its resources.\label{figure:pipeline}}
\end{figure}

Unlike traditional \smr{}s, \prsm leverages bandwidth, CPU and memory resources.
Traditional \smr{}s start a new consensus instance no sooner than when the previous consensus instance decides.
\prsm uses a different approach in order to leverage the network bandwidth and the CPU resources.
As we described in~\cref{ssec:pipeline-stage}, a pipeline epoch first receives transactions from clients to create a batch, then hashes this batch, broadcasts it and finally executes a consensus over small hash values.
This results in four distinct phases depicted in Figure~\ref{figure:pipeline}: a network reception phase (Rx), a CPU intensive hash phase (H), a network transmission phase (Tx) and a wait-mostly consensus phase (C).

The goal of \prsm is to have all these phases executing at the same time from different pipeline epochs, hence leveraging most resources.
To this end, a \prsm replica spawns a new pipeline epoch before the previous epoch terminates.
Figure~\ref{figure:pipeline} illustrates a 4-epoch pipeline where each replica of \prsm starts a new epoch as soon as its resources permit.
Each row stands for a pipeline epoch and has four phases, Rx, H, Tx and V.
As soon as four epochs run concurrently, the replica can execute all these distinct phases at the same time, one phase per epoch, and leverages most resources.
This technique called pipelining is especially effective when most of the transactions are issued by geodistributed clients to their closest replicas and when most concurrent transactions do not conflict because such phases consume different resources.
Recall that the fact that concurrent transactions are usually not conflicting was
observed before and benefited others~\cite{KAD07,MAK13}.

\section{Implementing the pipeline}\label{section:implementation}

In this section, we present how to implement a pipeline by making sure each replica can spawn a new pipeline epoch (\cref{ssec:phase-details}), how replicas naturally coordinate to participate in the same uniquely identified epoch (\cref{ssec:spawn-remote}), how the algorithm spawns new epochs (\cref{ssec:algo}), how batches are committed (\cref{ssec:pipeline-stage-order}), and how epochs are ordered (\cref{ssec:pipeline-order}).

\subsection{Distributed spawn of epochs}\label{ssec:phase-details}
The first phase (Rx) of a pipeline epoch (Fig.~\ref{figure:pipeline}) consists of receiving transactions from clients in the transaction pool, as we explained in~\cref{ssec:pipeline-stage}.
As a replica has no control over the number of transactions that the clients send, its only responsibility is to ensure that there is always space in the pool to collect new transactions.
If the pool is full, incoming client transactions are discarded and the client must retry later or on another replica.
This motivates the first condition to spawn a new pipeline epoch: having a full transaction pool.
Spawning a new epoch before the pool is full would result in fewer transactions per batch and thus in a lower throughput.
When the clients send transactions slowly, filling a transaction pool takes a long time.
To prevent old transactions from hanging in the transaction pool for too long, a replica also spawns a new pipeline epoch after a timeout expires, however, this never happens in our experiments as the demand is high.

The next phase (Tx) consists of broadcasting an epoch batch to the other replicas.
A pipeline epoch can only progress to its next phase if the previous epoch has finished the same phase.
A second condition for spawning a new epoch is thus to have an idle network, at least for the sending part.
To detect when the network is idle, each replica of \prsm continuously monitors its network usage.
Every 2\,ms, each node measures its sending rate and compares this sending rate after 3 samples (over the resulting 6\,ms duration) to the 600\,MiB/s physical network capacity (as we measure in~\cref{section:evaluation}).
If the network usage is lower than 5\% of the physical network bandwidth limit, then the network is considered idle.

It is also important for a node to spawn a new pipeline epoch only if it has enough memory available. 
The risk is, otherwise, for replicas to inflate the latency of a consensus instance by accepting to participate in too many concurrent epochs.
To decide the maximum number of epochs, each replica divides, prior to the execution, the amount of available memory (returned by the OS command \code{free}) by the batch size.
Each replica also keeps a few megabytes to store thread stacks, consensus instance states and other small sized objects.
This available memory is observed offline because it is hard to assess the memory available in real-time in Java due to the garbage collector of OpenJDK11 that is unpredictable.

\subsection{Following remote epochs}
\label{ssec:spawn-remote}

Each pipeline epoch corresponds to a new epoch in which one consensus instance executes.
By monitoring its own resource usage, a replica decides when to spawn a new pipeline epoch independently of the other replicas.
Each replica tags the pipeline epochs it spawns with an increasing \emph{epoch number} based on the current index of the pipeline---if two replicas spawn two pipeline epochs with the same number then it means that there are actually the same epoch.
Additionally, all the consensus messages associated with an epoch are prefixed by the epoch number of the epoch.
A replica considers that the received messages tagged with the same epoch number belong to the same pipeline epoch and thus to the same consensus instance.
With this simple method, replicas join an existing pipeline epoch by spawning an epoch locally which happens to have the same epoch number.

A replica participates to remotely spawned epochs depending on its local state and the received messages.
When every replica in the system receives an equal number of transactions from the clients, the replicas end up spawning new pipeline epochs at the same rate.
In this ideal scenario, all replicas participate to the same pipeline epochs without any additional synchronization.
When the clients load is imbalanced among the replicas, some replicas may never decide to spawn a new epoch on their own.
However, sufficiently many replicas must participate in a consensus instance for this instance to terminate.
For this reason, when a replica receives a message from a pipeline epoch to which it does not participate,  if the previous epoch is decided then this replica participates to the epoch.
This restriction of having the previous epoch decided prevents a malicious node from making other replicas participate in epochs in a distant future, which could lead to starvation.

\begin{algorithm}[t]
	\caption{\prsm at process $p_i$ \label{algo:dispel}
  }
  
  \renewcommand{\Comment}[1]{}
	{\footnotesize

	\begin{algorithmic}[1]
	
		\Part{Initially}
			\State $\ms{batch}_i \gets \emptyset$ {\color{gray}{\textit{// gathers the client requests}}}
			\State $\ms{stage-num}_i \gets 0$ {\color{gray}{\textit{// number of running instances}}}
                        \State $\ms{started}_i \gets \emptyset$ ; $\ms{decided}_i \gets \emptyset$ {\color{gray}{\textit{// epochs of started / decided instances}}}

			\State $\ms{max-stages} \gets \ms{available-memory}/\ms{batch-capacity}$ \label{line:mem}
		\EndPart
		
		\Statex	
			
%
		
		\Part{When $\ms{batch}_i$ is full, network is idle and $\ms{stage-num}_i < \ms{max-stages}$} \label{line:nw}
                  \State $\lit{spawn-instance}(\lit{max}(\ms{started}_i) + 1, \ms{batch}_i)$  {\color{gray}{\textit{// non-blocking call}}}
		  \State $\ms{batch}_i \gets \emptyset$
		\EndPart \label{line:limit-stages}

		\Statex

		\Part{When receive a message for $\lit{consensus}^{\ms{epoch}}$ and $epoch \notin \ms{started}_i$} \label{line:follow-start}
                \WUntil{$\lit{consensus}^{\ms{epoch}-1}$ decides}\EndWUntil \label{line:follow-condition}
                \If{$\ms{epoch} = \lit{max}(\ms{started}_i) + 1$} \label{line:follow-batch-start}
                \State $\lit{spawn-instance}(\ms{epoch}, \ms{batch}_i)$  {\color{gray}{\textit{// non-blocking call}}}
		\State $\ms{batch}_i \gets \emptyset$ \label{line:follow-batch-end}
                \Else{}
                $\lit{spawn-instance}(\ms{epoch}, \emptyset)$  {\color{gray}{\textit{// non-blocking call}}} \label{line:follow-empty}
                \EndIf
		\EndPart

%
		
		\Statex
		
		\Part{$\lit{spawn-instance}(\ms{epoch}, \ms{batch})$} {{\color{gray}{\textit{ // starts a new consensus instance}}}} \label{line:stage-start}
			\State $\ms{stage-num}_i \gets \ms{stage-num}_i+1$ ; $\ms{started}_i \gets \ms{started}_i \cup \{ \ms{epoch} \}$
			\State $\ms{results} \gets \lit{consensus}^{\ms{epoch}}(\ms{batch})$ {\color{gray}{\textit{// blocking call}}} \label{line:start-consensus}
			\State $\ms{decided}_i \gets \ms{decided}_i \cup \{ \ms{epoch} \}$
                        \If{$\ms{epoch} > 0$} \label{line:reorder-start}
                        \WUntil{$\lit{consensus}^{\ms{epoch}-1}$ decides}\EndWUntil
                        \EndIf
			\State transmit $\ms{results}$ to the application \label{line:reorder-end}
			\State $\ms{stage-num}_i \gets \ms{stage-num}_i-1$ \label{line:stage-end}
		\EndPart

	\end{algorithmic}
	}
\end{algorithm}

\subsection{The algorithm for spawning new pipeline epochs}\label{ssec:algo}

Algorithm~\ref{algo:dispel} summarizes with pseudocode how \prsm decides to spawn new epochs. 
If a replica $p_i$ meets the three conditions (a) the transaction pool $\ms{batch}_i$ is full, (b) its network is idle and (c) there is enough available memory, then the replica spawns a new pipeline epoch (lines~~\ref{line:nw}--\ref{line:limit-stages}).
The epoch number of this new epoch is the immediately greater integer than the epoch number 
of the newest pipeline epoch to which the replica $p_i$ participated.
Additionally, when a replica receives a message from a pipeline epoch to which it does not participate, it waits for the previous epoch to decide and then participates (lines~~\ref{line:follow-start},\ref{line:follow-condition}).
If this new epoch is the new epoch the replica $p_i$  has spawned on its own, it participates with the content of its transaction pool (lines~~\ref{line:follow-batch-start}--\ref{line:follow-batch-end}).
Otherwise, it participates with an empty batch (lines~~\ref{line:follow-empty}).
This avoids submitting a batch to a consensus instance that is likely to already have accepted $n-t$ batches from other replicas.
In this case, the batch submitted by the replica $p_i$ is not part of the decision set.
The epoch spawning routine (lines~~\ref{line:stage-start}--\ref{line:stage-end}) starts a new consensus instance (cf.~\cref{ssec:dbft}) tagged by an epoch number and delivers the consensus decisions in the order of the epoch numbers.
We detail how a replica orders the decision inside a epoch in~\cref{ssec:pipeline-stage-order} and across epochs in~\cref{ssec:pipeline-order}.

\subsection{Intra-epoch ordering between hashes and batches}
\label{ssec:pipeline-stage-order}
An interesting side effect of our pipeline is that correct replicas may decide upon the hash of batches before receiving the corresponding batches in full.
As we mentioned in~\cref{ssec:pipeline-stage}, because replicas decide locally to spawn a new epoch, the transactions reception and the batches exchange happen concurrently.
Additionally, as the consensus decides on hashes and not on batches, the consensus decision also happens concurrently with the batch exchange.
As a result of this concurrency, a replica sometimes decides on batches it has not yet received.
Indeed, as we described in~\cref{ssec:dbft}, the DBFT consensus algorithm starts by a reliable broadcast and then executes a binary consensus on every reliably delivered value.
During the first step of a reliable broadcast, a  \prsm correct replica sends the batch to all the replicas.

The two following steps of a reliable broadcast are all-to-all broadcasts used to prevent correct replicas
from delivering values broadcast by Byzantine to only a subset of the replicas or from delivering 
different values.
\prsm correct replicas only send hashes during these to all-to-all broadcasts in order to save bandwidth.
A reliable broadcast delivers its value when the replica delivers enough messages from the two all-to-all reliable broadcast.
As a result, a \prsm correct replica sometimes has the hash of a batch reliably delivered without knowing the corresponding batch.
This is not an issue since this is only possible if at least one correct replica knows the batch.
A replica handles this corner case by broadcasting a batch request for the decided hashes with an unknown batch at the end of each pipeline epoch.
When a replica knows all the decided batches, it sorts them with their hash as a sorting key.

\subsection{Inter-epoch ordering to guarantee starvation freedom}
\label{ssec:pipeline-order}
Having each pipeline epoch deciding an ordered set of batches is not sufficient.
Because of the concurrency between pipeline epochs, the order in which epochs are decided is not always the order in which they are spawned.
Replicas solve this issue by transmitting the pipeline epochs decisions sorted by the epoch number of these epochs.
More formally, a replica transmits the decision of a pipeline epoch with the epoch number $k$ to the application only after transmitting the decisions of the previous pipeline epochs for all $j < k$ to the application.
This corresponds to the lines~~\ref{line:reorder-start}--\ref{line:reorder-end} in  Algorithm~\ref{algo:dispel}.

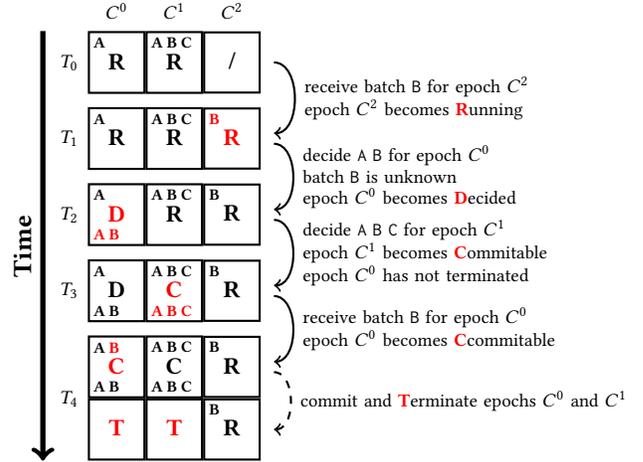
\begin{figure}[t]
  \begin{center}
    \begin{tikzpicture}
  \newcommand{\StageState}[3]
  {
    \begin{minipage}{17pt}
      \centering
          {\tiny{#1} \hfill}%
          \vspace{-3pt}
          \linebreak
          {\textbf{#2}}%
          \vspace{-4pt}
          \linebreak
          {\tiny{#3} \hfill}
    \end{minipage}
  }

  \tikzstyle{stage}=[draw, thick, minimum width=17pt, minimum height=17pt,
    font=\small\bf, inner sep=2pt]
  \tikzstyle{label}=[font=\scriptsize, text width=130pt]
  \tikzstyle{timepoint}=[font=\scriptsize]
  \tikzstyle{epoch}=[font=\scriptsize]
  \tikzstyle{time}=[draw,->,line width=2pt]
  \tikzstyle{event}=[->,thick]

  \node[stage] (s00) {\StageState{A}{R}{}};
  \node[stage, anchor=west] (s01) at (s00.east) {\StageState{A B C}{R}{}};
  \node[stage, anchor=west] (s02) at (s01.east) {\StageState{\color{white}N}{/}{}};
  \node[timepoint, anchor=east] (t0) at (s00.west) {$T_0$};

  \node[stage, anchor=north] (s10) at ($(s00.south) + (0, -5pt)$) {\StageState{A}{R}{}};
  \node[stage, anchor=west] (s11) at (s10.east) {\StageState{A B C}{R}{}};
  \node[stage, anchor=west] (s12) at (s11.east) {\StageState{\color{red}B}{\color{red}R}{}};
  \node[timepoint, anchor=east] (t1) at (s10.west) {$T_1$};

  \node[stage, anchor=north] (s20) at ($(s10.south) + (0, -5pt)$) {\StageState{A}{\color{red}D}{\color{red}A B}};
  \node[stage, anchor=west] (s21) at (s20.east) {\StageState{A B C}{R}{}};
  \node[stage, anchor=west] (s22) at (s21.east) {\StageState{B}{R}{}};
  \node[timepoint, anchor=east] (t2) at (s20.west) {$T_2$};

  \node[stage, anchor=north] (s30) at ($(s20.south) + (0, -5pt)$) {\StageState{A}{D}{A B}};
  \node[stage, anchor=west] (s31) at (s30.east) {\StageState{A B C}{\color{red}C}{\color{red}A B C}};
  \node[stage, anchor=west] (s32) at (s31.east) {\StageState{B}{R}{}};
  \node[timepoint, anchor=east] (t3) at (s30.west) {$T_3$};

  \node[stage, anchor=north] (s4a0) at ($(s30.south) + (0, -5pt)$) {\StageState{A \color{red}B}{\color{red}{C}}{A B}};
  \node[stage, anchor=west] (s4a1) at (s4a0.east) {\StageState{A B C}{C}{A B C}};
  \node[stage, anchor=west] (s4a2) at (s4a1.east) {\StageState{B}{R}{}};

  \node[stage, anchor=north] (s4b0) at (s4a0.south) {\StageState{\color{white}N}{\color{red}T}{}};
  \node[stage, anchor=west] (s4b1) at (s4b0.east) {\StageState{\color{white}N}{\color{red}T}{}};
  \node[stage, anchor=west] (s4b2) at (s4b1.east) {\StageState{B}{R}{}};

  \node[timepoint, anchor=east] (t4) at (s4a0.south west) {$T_4$};

  \node (timestart) at ($(s00.north west) + (-17pt, 0)$) {};
  \node (timeend) at ($(s4b0.south west) + (-17pt, 0)$) {};
  \draw[time] (timestart.center) -- node[anchor=south, rotate=90, font=\bf] {Time} (timeend.center);

  \foreach \i in {0,...,2} {
    \node[epoch, anchor=south] (e\i) at ($(s0\i.north) + (0, 1pt)$) {$C^\i$};
  }

  \draw[event] ($(s02.east) + (5pt, 0)$) to [out=0,in=0] node[label, anchor=west] {receive batch \texttt{B} for epoch $C^2$\\epoch $C^2$ becomes \textbf{\color{red}R}unning} ($(s12.east) + (5pt, 2pt)$);

  \draw[event] ($(s12.east) + (5pt, -2pt)$) to [out=0,in=0] node[label, anchor=west] {decide \texttt{A B} for epoch $C^0$\\batch \texttt{B} is unknown\\epoch $C^0$ becomes \textbf{\color{red}D}ecided} ($(s22.east) + (5pt, 2pt)$);

  \draw[event] ($(s22.east) + (5pt, -2pt)$) to [out=0,in=0] node[label, anchor=west] {decide \texttt{A B C} for epoch $C^1$\\epoch $C^1$ becomes \textbf{\color{red}C}ommitable\\epoch $C^0$ has not terminated} ($(s32.east) + (5pt, 2pt)$);

  \draw[event] ($(s32.east) + (5pt, -2pt)$) to [out=0,in=0] node[label, anchor=west] {receive batch \texttt{B} for epoch $C^0$\\epoch $C^0$ becomes \textbf{\color{red}C}commitable} ($(s4a2.east) + (5pt, 2pt)$);

  \draw[event, dashed] ($(s4a2.east) + (5pt, -2pt)$) to [out=0,in=0] node[label, anchor=west] {commit and \textbf{\color{red}T}erminate epochs $C^0$ and $C^1$} ($(s4b2.east) + (5pt, 0)$);
\end{tikzpicture}
  \end{center}
  \caption{\label{figure:outoforder}The consensus reordering pipeline of \prsm 
    indicates a possible evolution of pipeline epochs ($C^0$ to $C^2$) as events occur.
    The character in the middle of each cell indicates the epoch state, either running (R), decided (D), ready for commit (C) or not yet started (/).
    The top left characters indicate the received batches \texttt{A}, \texttt{B} or \texttt{C}.
    The bottom left characters indicate the batches the decided hashes stand for.}
\end{figure}

Figure~\ref{figure:outoforder} illustrates how both intra and inter-epoch orderings take place.
It shows the evolution of three pipeline epochs on a replica when this replica receives messages.
Initially ($T_0$), only two epochs $C^0$ and $C^1$ are running.
The replica has received the batch \texttt{A} for $C^0$ and the batches \texttt{A}, \texttt{B} and \texttt{C} for $C^1$.
Upon reception of the batch \texttt{B} for $C^2$ from another replica ($T_1$), the receiving replica spawns locally the new pipeline epoch $C^2$ as described in lines~~\ref{line:follow-start}--\ref{line:follow-batch-end} of Algorithm~\ref{algo:dispel}.
Upon reception of the consensus messages for $C^0$ ($T_2$), the replica reaches consensus and decides on the batches \texttt{A} and \texttt{B}.
However, as we explain in \cref{ssec:pipeline-stage-order}, the replica has not received the decided batch \texttt{B}.
Upon reception of this batch, the epoch $C^0$ is marked as decided but not yet ready to \emph{commit} (i.e., to transmit the decision to the application).
The replica then receives consensus messages for $C^1$ ($T_3$) and decides on the batches \texttt{A}, \texttt{B} and \texttt{C}.
Since the replica knows the decided hashes, it is ready to commit $C^1$ but waits for the previous epoch $C^0$ before transmitting to the application.
Finally, the replica receives the batch \texttt{B} for $C^0$ ($T_4$).
The replica then commits both epochs $C^0$ and $C^1$ and mark them as terminated.

To conclude, we summarize three properties for the distributed pipelining that allows to proceed efficiently:
\begin{enumerate}
\item \underline{Early epoch}: this requires launching a epoch $k+1$ early, typically epoch $k+1$ should start before the epoch $k$ completed in order to guarantee that the pipeline translates into higher throughput.
\item \underline{Oldest epoch priority}: epoch $k$ always has priority over epoch $k+1$
to ensure that 
the number of concurrently active epochs eventually decreases, 
preventing undesirable situations like starvation.
\item \underline{Pipeline feeding}: the \smr must guarantee that commands are enqueued into the upcoming epochs to guarantee progress, despite the workload induced by the inter replicas communications.
\end{enumerate}
Guaranteeing these three properties help ensure the good performance of the distributed pipelining. 

\section{\smr Evaluation}
\label{section:evaluation}

This section presents an evaluation of \prsm in geo-distributed setups with up to 256 machines.
We compare the performance of \prsm against four \smr{}s protocols from the literature. The larger experiments with a cryptocurrency application are deferred to~\cref{sec:blockchain}.

\subsection{\smr{}s protocols}
For this evaluation, we implement \prsm, a leaderless \smr using the DBFT consensus described in \cref{section:background} and the pipelined architecture described in \cref{section:design} and \cref{section:implementation}.
\prsm is written in Java using only the standard Java libraries.
We compare \prsm against four \smr{}s:
\begin{itemize}
\item \underline{\bftsmart}: is a leader-based BFT \smr similar to PBFT with further optimizations that has been maintained in Java for more than a decade~\cite{BSA14}.
  It was used for the ordering service of Hyperledger Fabric~\cite{ABB18} to tolerate Byzantine failures~\cite{SBV18}.
  We used the branch \code{weat2} of the official \bftsmart git repo \url{https://github.com/bft-smart/library} that outperforms the master branch when geo-distributed~\cite{SB15,BRS19}.
\item \underline{\epaxos}: is a leaderless {\bf crash fault tolerant} (CFT) \smr~\cite{MAK13} that does not provide the guarantees of a BFT \smr but that serves as a fast baseline. 
  It is an improved version of \paxos with no leader and a fast path for non conflicting commands.
  %
  We use the Go author's library from \url{https://github.com/efficient/epaxos.git}.
\item \underline{\hotstuff}: is a recent leader-based BFT \smr.
  It outperforms \bftsmart by having the leader piggybacking phases of distinct consensus instances into the same messages~\cite{YMR19}.
  Facebook is currently developing in Rust the Libra state machine~\cite{BBC19} on a variant of HotStuff~\cite{Fac19}.
  We use the C++ authors' library from \url{https://github.com/hot-stuff/libhotstuff}.
\item \underline{\zyzzyva}: is a leader-based BFT \smr designed to reduce decisions latency~\cite{KAD07}.
  To this end, \zyzzyva replicas  reply optimistically to the client before reaching consensus and let the client solve disagreements.
  We use the Zlight C++ implementation~\cite{AGK14} that we patched to make it run on a geo-distributed setup.
  More precisely, the original implementation uses IP multicast and UDP communication without any mechanism to handle packet loss.
  We replace the UDP implementation by a standard event-driven loop TCP implementation.
\end{itemize}
\underline{Other SMRs:} there is a long body of BFT \smr{}s papers, however, not all implementations are available: we contacted the authors of \bftmencius~\cite{MBS13} and \ezbft~\cite{APR19}, two leaderless BFT protocols, but there were no readily available implementation.

\subsection{Benchmark setup}\label{ssec:bsetup}
This evaluation takes place on AWS EC2 instances.
We use \textit{c5.xlarge} instances for running replica and client processes.
Each instance is a KVM virtual machine with four hardware threads implemented by two hyperthreaded Intel Xeon core running at 3 GHz, with 8 GiB of memory and a network interface of 600 MiB/s upload and download speed.
Each virtual machine is running a Ubuntu 18.04 distribution with an AWS modified Linux kernel 4.15, OpenJRE 11 and the glibc 2.27.
The \smr{}s are compiled using OpenJDK 11, gcc 7.5 and go 1.10.4.

We ran  \bftsmart, \zyzzyva, \hotstuff and \epaxos in their default configurations indicated in their 
original papers or reports~\cite{BSA14,AGK14,YMR19,MAK13}.
Each \prsm replica uses a batch size of $25/n$ MB where $n$ is the number of replica.
We configure \prsm replicas to use up to 12 pipeline epochs.
In every experiment, we execute one replica process per VM.
In addition, we dedicate two VMs per region to execute the client processes.
For \epaxos, \bftsmart, \hotstuff and \prsm, each client VM executes a single client process which sends many parallel transactions to the replicas.
\zyzzyva clients only implement blocking requests.
For \zyzzyva, we execute 100 client processes per client VM.
We found that this number of client process is sufficient to saturate \zyzzyva replicas.

The goal of the following experiments is to compare the throughput that each \smr delivers.
This throughput increases with the client load.
As a tradeoff, increasing the client load generally results in a higher latency.
To provide a fair comparison, we choose to explore different client loads for every data points we plot.
We then select the client loads resulting in a throughput of at least 90\% of the best observed throughput.
Among these selected loads, we pick the one resulting in the best latency.
The intuition behind this methodology is that when the client load increases, the latency remains stable until an inflection point and then skyrockets.
The throughput increases until this inflection point and then remains stable.
Our methodogy identifies the inflection point by selecting the set of client loads with the best throughputs and then pick the latency obtained in this set right before the inflection point.
All the data points are the average over at least 5 runs.

\begin{table}[t]
\setlength{\tabcolsep}{8pt}
  \begin{tabular}{|r|cccc|}
    \cline{2-5}
    \multicolumn{1}{c|}{} & Frankfurt & Ireland & London & Paris \\
   \hline
    Frankfurt & --        & 60      & 109    & 179 \\
    Ireland   & 25        & --      & 148    & 80  \\
    London    & 14        & 10      & --     & 210 \\
    Paris     & 8         & 19      & 7      & --  \\
    \hline
  \end{tabular}
  \caption{\label{table:setup-eu}Latency in ms (bottom left) and bandwidth in MiB/s (top right) between the AWS Europeans datacenters.}
\end{table}

\subsection{Pipelining copes with round-trip delays}\label{sec:throughput}

As we expect the pipeline to increase throughput despite high latency, we start by comparing the throughput and latency of \prsm to \epaxos, \hotstuff, \bftsmart and \zyzzyva within a \emph{single} continent.
More precisely, we measured the throughput and latency on up to 256 VMs located in Europe (Ireland, London, Paris and Frankfurt).
Table~\ref{table:setup-eu} summarizes the latencies and bandwidth between the four datacenters, mesured with \texttt{nuTCP}.


%
%
\begin{figure*}[t]
  \centering
  \subfigure[\label{fig:scale-throughput}Throughput each \smr.]{\includegraphics[keepaspectratio=true,width=.51\textwidth]{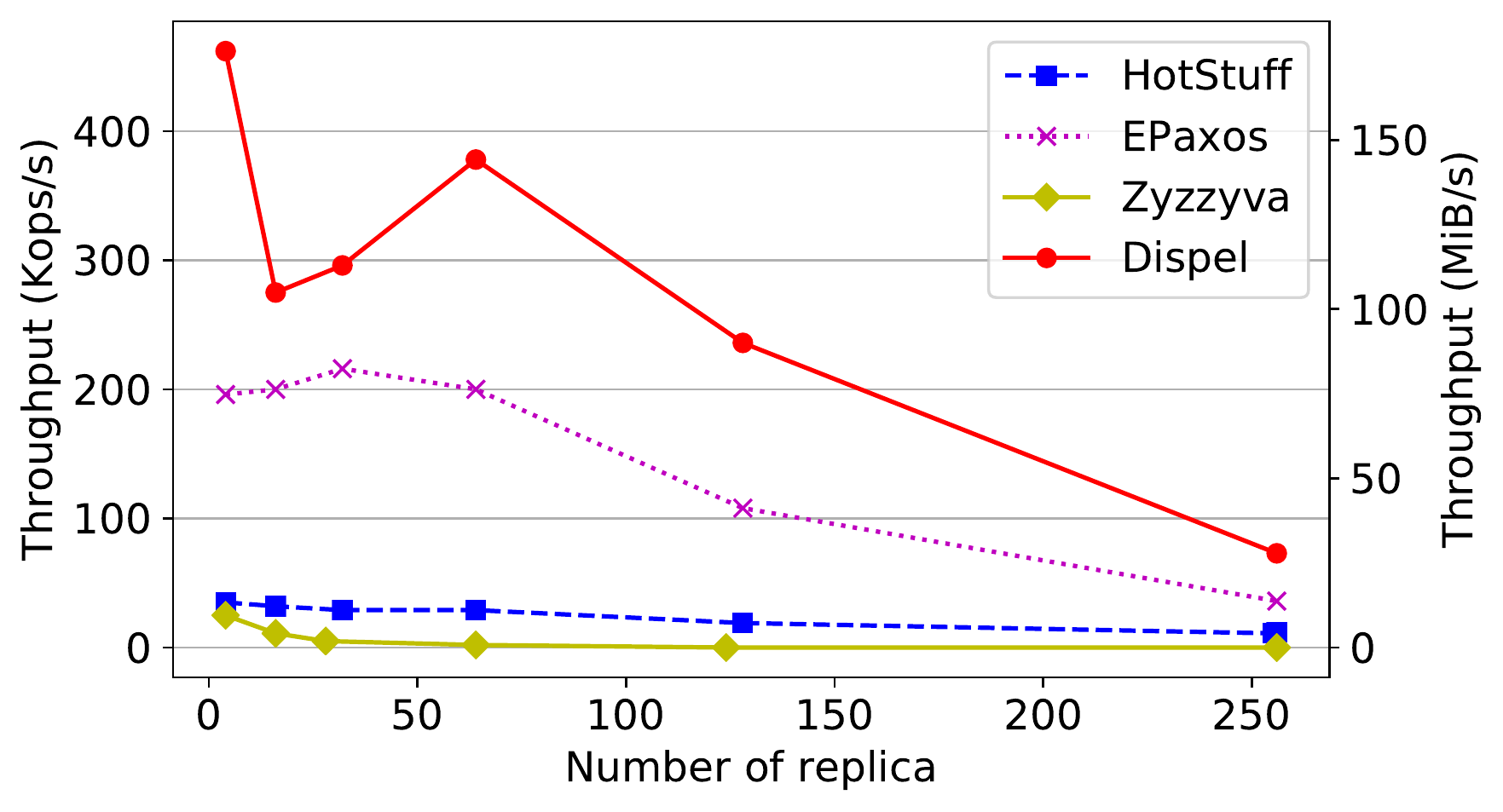}}\quad
  \subfigure[\label{fig:scale-latency}Latency of each \smr (y-axis in logscale).]{\includegraphics[keepaspectratio=true,width=.47\textwidth]{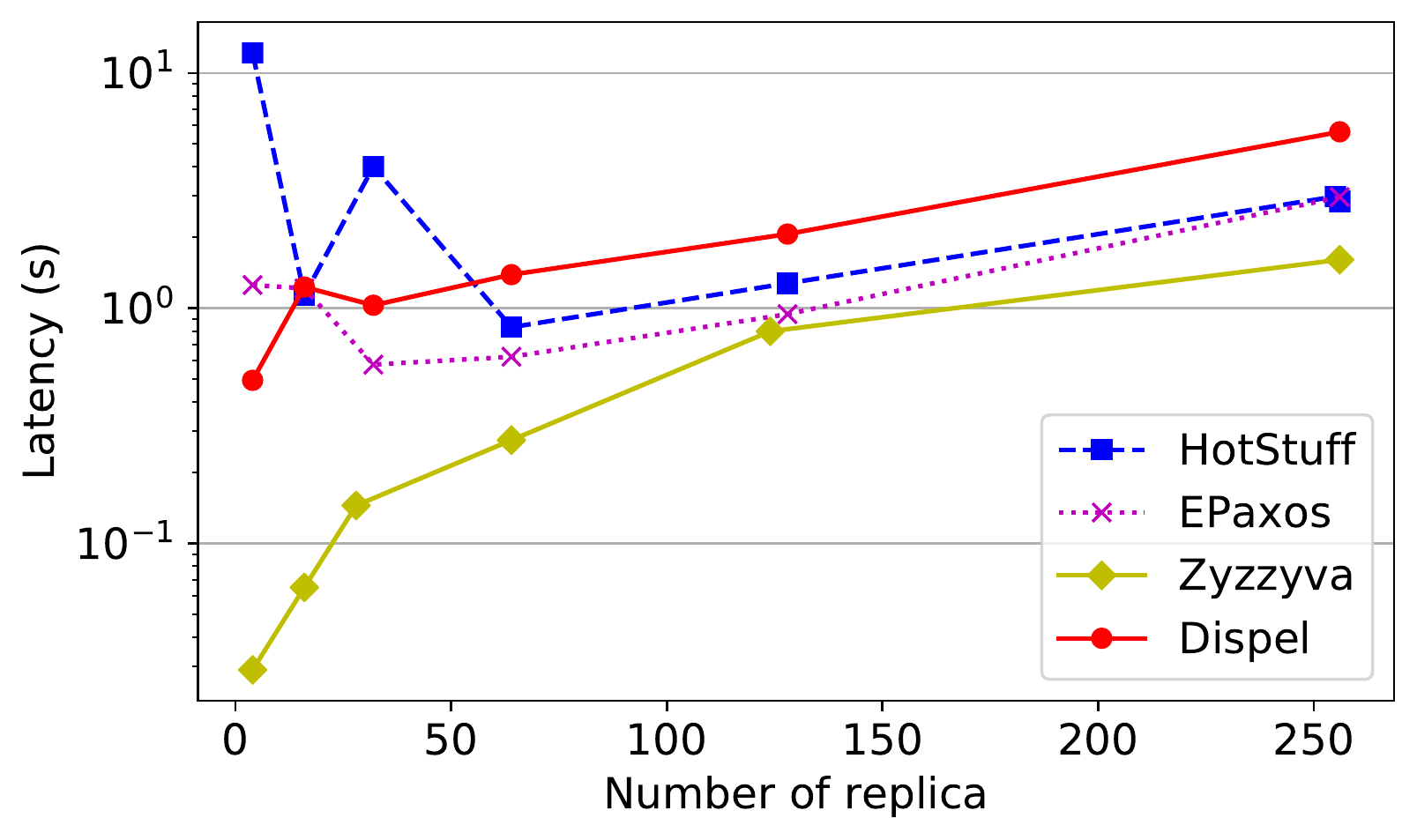}}
  \caption{\label{figure:scale}Performance of \prsm, \hotstuff, \epaxos and \zyzzyva depending on the number of nodes spread in 4 European datacenters (Ireland, London, Paris and Frankfurt) with an increasing number of nodes committing 400-bytes transactions.}
\ifnottechrep\vspace{-1em}\fi
\end{figure*}


Figure~\ref{fig:scale-throughput} depicts the throughput of \prsm, \epaxos, \hotstuff and \zyzzyva.
We found that \bftsmart and \hotstuff exhibit the same behavior although \bftsmart throughput is consistently at least $2\times$ lower than \hotstuff throughput.
We choose to not plot \bftsmart for sake of clarity.
First, we observe that \zyzzyva does not perform as well as the others with a throughput of 25 Kops/s at 4 nodes and of 2 Kops/s or below with 64 nodes or more.
This is because \zyzzyva is optimized to perform in a local area network (LAN): it does not batch requests and executes sequentially
Second, as expected, \epaxos outperforms \hotstuff by delivering a throughput up to $7\times$ larger at 32 nodes and still $3\times$ larger at 256 nodes.
\epaxos does not offer the same level of fault tolerance than \hotstuff: it does not tolerate Byzantine failures while \hotstuff does. 
Third, all solutions perform significantly slower than \prsm due to their lack of pipeline.
\prsm exhibits the best performance at 4 nodes with a throughput of 462 Kops/s (more than $2\times$ larger than \epaxos and $13\times$ larger than \hotstuff).
At worst, with 256 nodes, \prsm has a throughput of 73 Kops/s (still $2\times$ larger than \epaxos and $6\times$ larger than \hotstuff).

Pipelining typically hides the increase in latency through parallelization, executing multiple consensus instances at a time.
Although \hotstuff pipelines to some extent, its pipeline only piggybacks phases of consecutive consensus instances into the same messages and does not leverage resources from any replicas like \prsm's pipeline does.
%
%
%
Interestingly, \prsm also doubles the performance of \epaxos at 4, 128 and 256 nodes even though \epaxos does not tolerate Byzantine failures.
As we consider only non-conflicting requests here, all command leaders of \epaxos should commit their command in parallel---similarly to a pipeline. The difference is that EPaxos does not spawn new leader proposal based on resources as \prsm does, hence unable to detect resource contention. 

A surprising phenomenon is the unexpected high throughput of \prsm with 4 replicas and the througput drop between 4 and 16 replicas.
As we describe later in \cref{ssec:hol}, we expect the throughput of \prsm to increase with the number of replicas while the number of replica is small, then to decrease.
This is what we observe from 16 replicas, however the throughput of \prsm at 4 replicas is $1.68\times$ larger than at 16 nodes.
Although we do not have a definitive explanation for this throughput, we are confident that it does not come from either our measurement nor the methodology we use.
Indeed, over the 5 runs we use to compute the average throughput of this point, the individual throughputs are all between 449 Kops/s and 499 Kops/s.
The client load we use is the one that results in the best throughput over the explored values.
Our main hypothesis is that for this small number of replica, all the replicas progress at the same rate in the pipeline, which is the ideal scenario that we describe in \cref{ssec:spawn-remote} where all replicas always participate with full batches.


\subsection{Higher performance than SMR for blockchains}\label{sec:latency}
Figure~\ref{fig:scale-latency} depicts the latencies obtained for the four \smr{}s, \prsm, \epaxos, \hotstuff and \zyzzyva, in the same settings as above (\cref{sec:throughput}). This
shows that \prsm{} does not offer the best latency. This is because the 
pipeline does not need to reduce the latency to increase throughput. Zyzzyva, designed for LANs, offers consistently the lowest latency by staying below 300 ms until 64 nodes and always below 1700 ms.
In contrast with other solutions, the latency of HotStuff improves with the system size at small scale, which is likely due to our measurements that extract the best throughput runs of all runs before selecting the best latency run within these values (\cref{ssec:bsetup}).
Due to HotStuff lower communication complexity, HotStuff latency increases slower than other \smr{}s.
As a result, the best throughput values are observed in conditions where all latencies are relatively similar.
%
Given that Facebook aims at deploying a variant of HotStuff on 100 nodes and more as part of their Libra blockchain~\cite[\S{}5]{Fac19}, it is particularly interesting to compare performance at 128 nodes: the latencies of \hotstuff and \prsm are in the same order of magnitude, respectively 1300 ms and 2100 ms, whereas the throughput of \prsm{} (Fig.~\ref{fig:scale-throughput}) is more than $12\times$ higher than HotStuff. We realized that the default HotStuff is limited so in~\cref{sec:blockchain} we tune HotStuff to obtained better performance on geodistributed experiments, however,  as we will see, it remains significantly slower than \prsm.

\begin{table}[t]
\setlength{\tabcolsep}{17pt}
\begin{center}
  \begin{tabular}{|c|c|c|}
  \cline{2-3}
  \multicolumn{1}{c|}{} & \multicolumn{2}{c|}{Throughput (MiB/s)} \\
  \hline
  Parallel streams & Total & Per stream \\
  \hline
  1 & 4 & 4.53 \\
  10 & 45 & 4.53 \\
  50 & 225 & 4.51 \\
  100 & 443 & 4.43 \\
  200 & 533 & 2.67 \\
  \hline
\end{tabular}

\end{center}
\caption{\label{table:tcphol}Bandwidth between two machines located in Sydney and S\~{a}o Paulo
  depending on the number of parallel TCP streams obtained by \code{nutTCP} and showing that HOL blocking limits performance. 
}
\end{table}

\subsection{Multiplying TCP bandwidth capacity}\label{ssec:hol}

Because TCP preserves ordering, using one connection to transmit independent data often leads to the head-of-line (HOL) blocking suboptimal phenomenon~\cite{SK06} where the loss of one packet actually delays the reception of potentially numerous subsequent packets until after the packet is retransmitted and successfully delivered.
As simple way to circumvent this phenomenon is to use several TCP connections in parallel.
Table~\ref{table:tcphol} illustrates the difference between the NIC physical limit and the bandwidth limit of a single TCP connection. 
As one can see, the network usage increases linearly with the number of parallel connections until the limit of the network interface is met at around 533\,MB/s, after which the bandwidth per stream decreases due to this physical limit.

Despite having several open TCP connections, one per peer, sequential \smr{}s replicas are unable to benefit from this parallelism.
Indeed, when such sequential replicas broadcast a message, they must wait the transmission of this message completes on all the TCP connections before to progress.
On the contrary, \prsm replicas do not wait the transmission completion of every TCP connection thanks to their pipeline structure.

\begin{figure}
\begin{center}
  \includegraphics[keepaspectratio=true, width=.48\textwidth]{%
    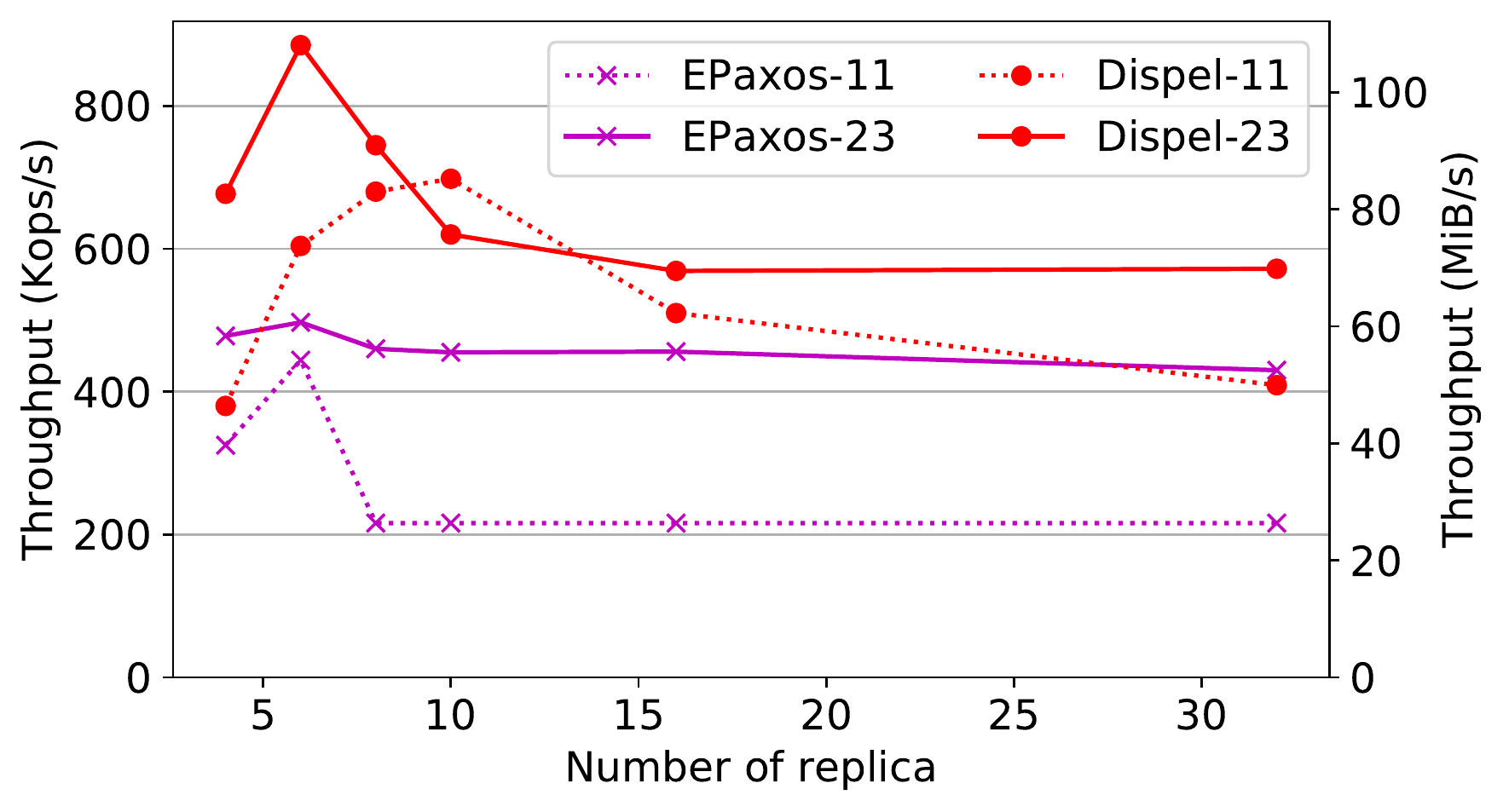}
\end{center}
\caption{\label{figure:netparam-throughput} Effect of HOL blocking on the throughput of \prsm and \epaxos with an increasing number of nodes committing 128-bytes transactions for a TCP-bandwidth of 23\,MiB/s between North California and North Virginia, and of 11\,MiB/s between North Virginia and Tokyo.}
\end{figure}


To illustrate how \prsm cirumvents the HOL blocking phenomenon, we measure the throughput of \epaxos, \hotstuff, \zyzzyva and \prsm when the number of replica varies on two setups.
The first setup evenly spreads the replicas across two regions, North Virginia and North California.
The bandwidth of one TCP connection between these two regions if 23\,MiB/s.
The second setup evenly spreads the replicas across North Virginia and Tokyo.
The TCP bandwidth between these two regions 11\,MiB/s.
On these two setups where the TCP bandwidth is low, \prsm benefits from an increasing number of replicas.
However, when the number of replicas becomes too large, the probability of having a packet loss on a majority of TCP connections becomes significant and \prsm does not benefit from additional replicas.

We report on Figure~\ref{figure:netparam-throughput} the throughput for \prsm and \epaxos on the two setups, 11\,MiB/s and 23\,MiB/s of TCP bandwidth, when the number of replica increases with transactions of 128 bytes.
We do not report the throughput for \hotstuff, \bftsmart and \zyzzyva as their throughput is systematically below respectively 20 Kops/s, 14 Kops/s and 6 Kops/s.
We observe that for each setup and every number of replica, the throughput of \prsm is larger than for the other \smr{}s.
Moreover, the throughput of \prsm first increases up to 108\,MiB/s at 6 replicas in the 23\,MiB/s bandwidth setup and up to 85\,MiB/s at 10 replicas in the 11\,MiB/s bandwidth setup.
This increase is caused by \prsm replicas using more TCP connections in parallel when the number of replicas increases.
We observe a similar yet less pronounced evolution for \epaxos.
Indeed, \epaxos replicas decide transactions in parallel as long as they are non conflicting.
This parallelism makes \epaxos also using TCP connections in parallel although to a lesser extent.
While we do not observe this phenomenon on \hotstuff, \bftsmart and \zyzzyva, their throughput is too small to conclude.

\subsection{Robustness in case of failures}\label{ssec:robust}

In order to assess the performance of \prsm and \hotstuff under failures, we ran an experiment for 122 seconds on $n=32$ nodes spread evenly in Ireland, London, Paris and Frankfurt during which we manually injected crash failures. In this experiment, each transaction is of size 400 bytes, similar to the size of simple Bitcoin transactions.

\paragraph{Boosting HotStuff performance for geo-distribution.}
As we observed in~\cref{sec:latency}, HotStuff performance is particularly limited by default. As far as we know, HotStuff has only been evaluated in a single datacenter where the ``geodistributed'' experiments were emulated by adding artificial latencies~\cite{Agm18}.
As we show in \cref{ssec:hol}, real geodistributed setups, where large blockchains are expected to be deployed, also suffer from packet loss forcing TCP connections to block during packet retransmissions.
As an example, during the Red Belly Blockchain experiments~\cite{CNG18}, the bandwidth of a TCP connection between the AWS availability zones of Sydney and S\~{a}o Paulo was only 5\,MB/s.
To cope with this limitation, we increased the size of the batches used by \hotstuff to obtain the same decision size of 25\,MB in both \hotstuff and \prsm in large-scale experiments.
When \hotstuff uses larger batches, there are proportionally more packet loss during the batch transmission than during the other phases of consensus.
Since a replica broadcasts batches to every peers in parallel, larger batches brings more opportunities for parallel TCP transmissions.
While this increases the throughput, it comes at the cost or larger latencies.
In our case, using 25\,MB batches results in an average request latency of 20 seconds.


\begin{figure*}[t]
\begin{center}
\includegraphics[keepaspectratio=true, width=1\textwidth]{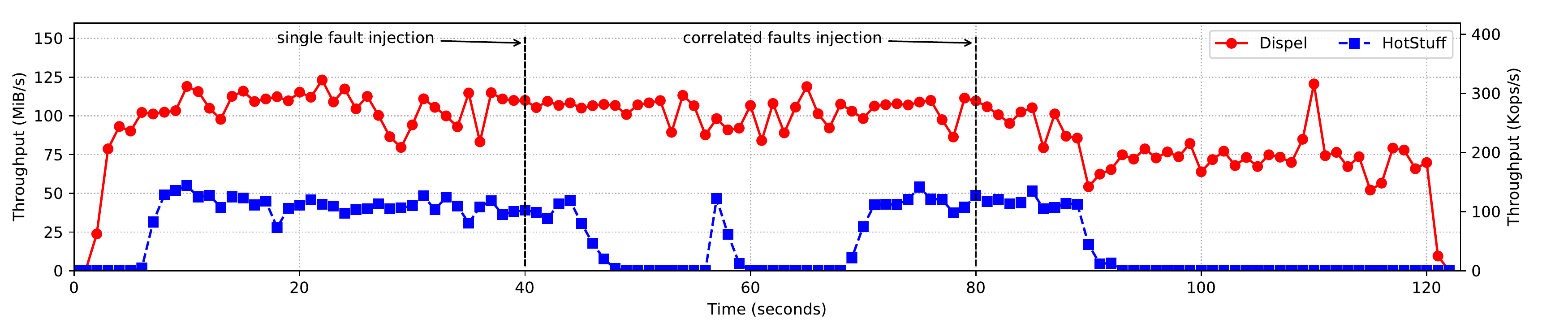}
\end{center}
\caption{\label{figure:failure-goodput}Robustness to failures as the throughput and evolution of the transactions per second over time for \hotstuff and \prsm. In a network of 32 nodes, we inject a single fault at 40 seconds in a region and additional correlated failures affecting all the remaining 7 nodes of this region at 80 seconds. 
}
\end{figure*}


\paragraph{Impact of isolated and correlated failures.}
Figure~\ref{figure:failure-goodput} depicts the performance expressed as the throughput in MiB/second and the number of transactions committed per second by the two protocols. First, one can see that \hotstuff takes more time than \prsm to reach its peak throughput that is lower than \prsm. 
These two observations confirm 
previous conclusions 
that the leader-based pattern leads to higher latency and lower throughput~\cite{CNG18}. 

More interestingly, we manually injected a failure of the leader on \hotstuff and the weak coordinator of the binary consensus on \prsm. As \hotstuff triggers a view-change when the faulty leader is detected as slow, the system must wait for the detection to occur and for the view-change to complete before the throughput can increase again. Again this phenomenon was already experienced in leader-based \smr{}s, like Zookeeper~\cite[Fig.6]{SRM12}, Multi-Paxos~\cite[Fig.10]{MAK13}, Mir-BFT~\cite[Fig.8]{SDM19} and Paxos~\cite[Fig.8]{EBR20}.  More surprisingly, a second view-change seems to occur systematically after one failure, this can be seen at 57 seconds. As no single node correctness is required for termination in \prsm{}, the throughput does not seem to drop  in \prsm after a single failure.

We also injected correlated failures to see the performance variations in \prsm and \hotstuff.
The correlated failures consists of shutting down all the remaining machines of the Frankfurt region.
The throughput of \hotstuff drops to 0 whereas the throughput of \prsm{} drops by about 30\%.
Note that the impact of failures on performance of leader-based \smr{}s raises the question of their suitability for cryptocurrency applications where simply DDoS-ing one node, the leader, or a region can DDoS the entire system.

\section{Cryptocurrency Application}\label{sec:blockchain}

We now present the performance of \prsm within a cryptocurrency application on up to 380 machines deployed over 3 continents and show that the bottleneck is not \prsm but the cryptographic verifications.

\begin{figure}[t]
\begin{center}
  \includegraphics[clip=true,viewport=3 10 487 205, keepaspectratio=true, width=.48\textwidth]{%
    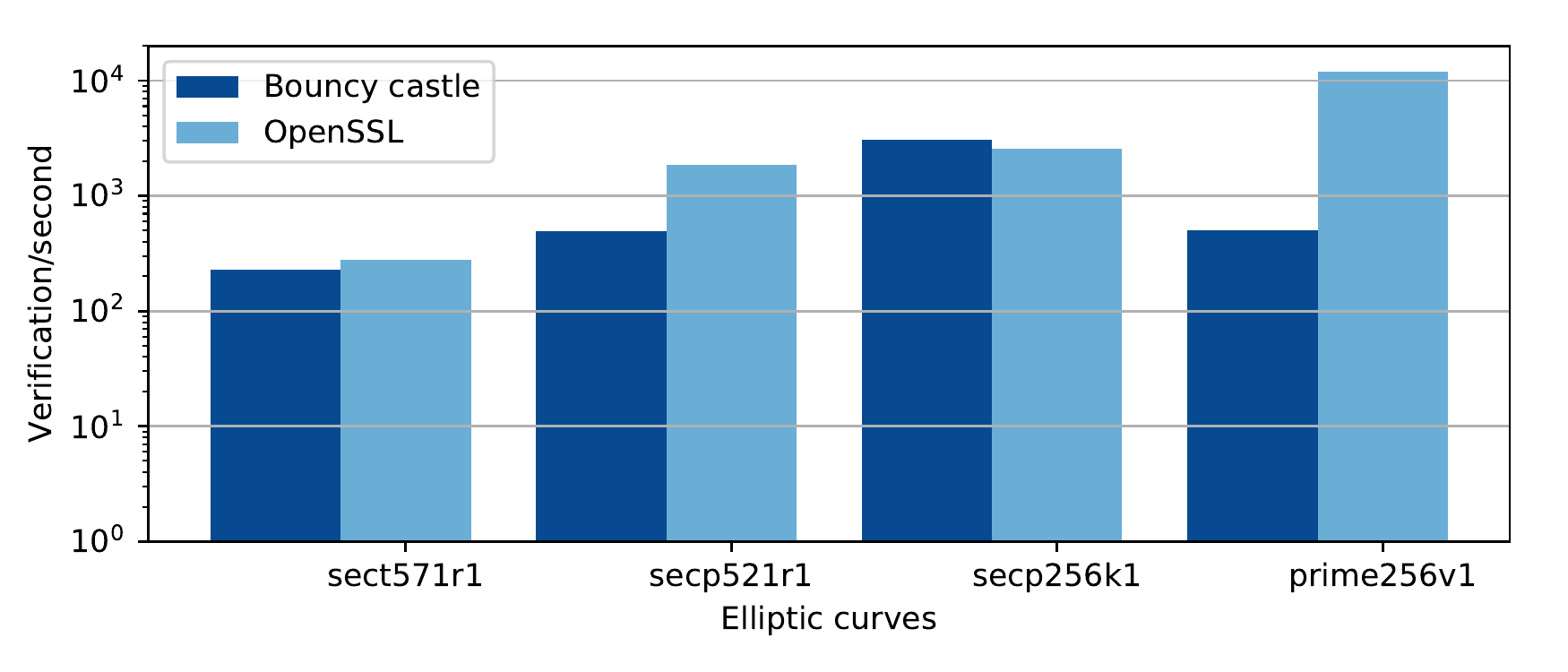}
\end{center}
\caption{\label{figure:crypto}Verification per second on one core for various
  elliptic curves. Note the logarithmic scale on the y-axis.}
\end{figure}

\paragraph{Application cryptographic overhead.}
To understand if an \smr can be used for blockchain applications like a cryptocurrency where assets are transferred among owners of some accounts represented with a public key, one must evaluate the performance of the \smr when transactions are cryptographically signed by the clients and these signatures are verified by the replica. 
Figure~\ref{figure:crypto} depicts the time it takes to verify signatures with one core using the public-key cryptosystems available in OpenSSL and written in C, and in Bouncy Castle and written in Java. One can observe the large variation of performance of the cryptosystems and the overhead of the Java library.


\paragraph{\prsm with a cryptocurrency application.}
We extended \prsm to support a cryptocurrency application by signing and verifying all transactions. Each user is equipped with an account and a public-private key pair. Each client pre-generates signed transactions using its private key prior to sending them to some replica, whereas every correct replica that receives a transaction verifies it using the public key associated with the account. Once the verification is correct the transaction can be stored.

\paragraph{From blockchain to block-sequence.}
Note that the hash-link between blocks is not necessary under the assumption that $f<\frac{n}{3}$ processes can fail as one can simply retrieve an immutable copy of a block by requesting it from $2f+1$ correct processes. By definition, among these responses, all correct replicas will respond with the same tamper-proof copy. Hence one has to find the copy duplicated
$f+1$ times among the $2f+1$ ones to identify it as the correct copy of the block. 

\paragraph{Performance at large-scale.}
For this experiment, we used up to 380 \textit{c5.2xlarge} replica VMs located equally in 10 regions on 3 continents: California, Canada, Frankfurt, Ireland, London, North Virginia, Ohio, Oregon, Sydney and Tokyo.
The \textit{c5.2xlarge} VMs have the same configuration than the \textit{c5.xlarge} VMs we use in \cref{section:evaluation} but with 8 hardware threads instead of 4 and 16 GiB of memory instead of 8 GiB.
For sending transactions, we spawned 10 additional client VMs, one from each region.
We selected the per-node batch size to be 2\,MB for 10 replicas, 8\,MB for 40 replicas and 24\,MB for 120, 250, 380 replicas to minimize the duration of the experiments. We use the \code{secp256k1} public-key cryptosystem that is used by Bitcoin~\cite{Nak08} and transactions of size 400 bytes, just like Bitcoin UTXO transactions.

\begin{figure}[t]
\begin{center}
  \includegraphics[keepaspectratio=true, clip=true,viewport= 0 10 495 165, width=.48\textwidth]{%
    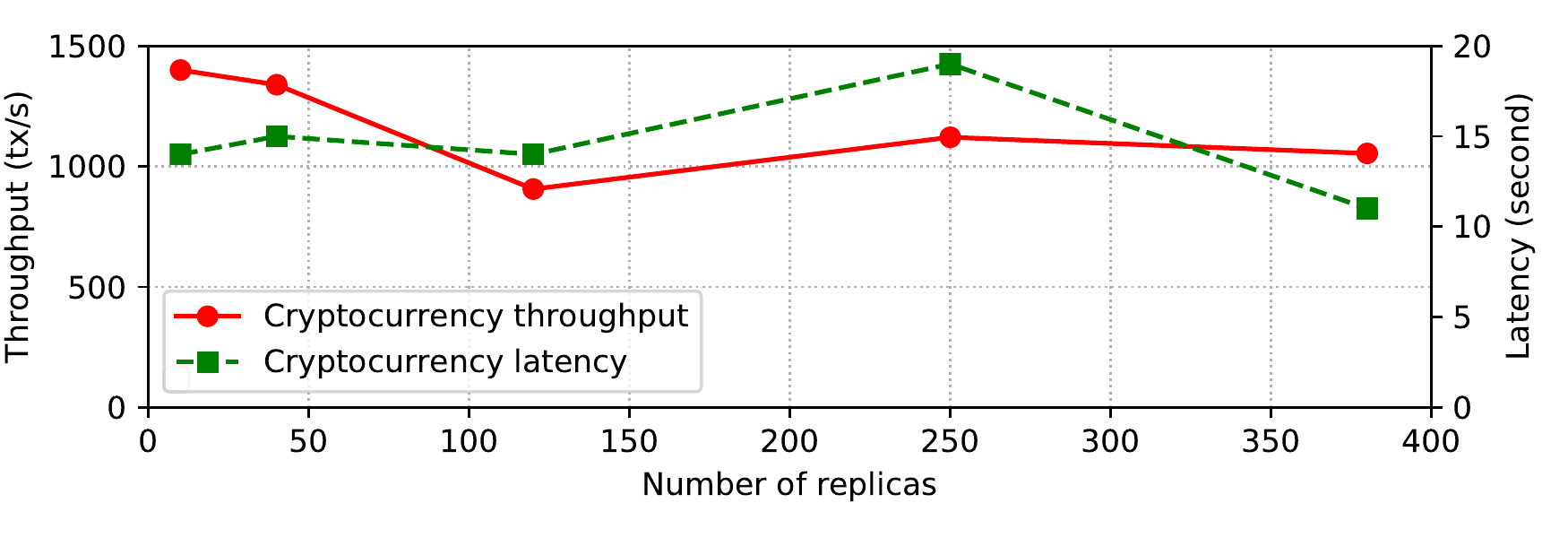}
\end{center}
\caption{\label{figure:cryptocurrency}Performance of a cryptocurrency running on top of \prsm as the system size increases from 10 to 380 nodes equally balanced across 10 datacenters over 3 continents.}
\end{figure}

Figure~\ref{figure:cryptocurrency} depicts the performance of \prsm when executing a cryptocurrency application where all transactions are verified by all replicas. We represent the transactions committed per second, including the signatures in addition to the application payload that contains the simple transfer information.  One can see that the throughput and latency vary slightly across the different network sizes but there is no clear difference between the performance on 10 nodes or 380 nodes, indicating that \prsm{} is not the bottleneck when running the cryptocurrency application. 
In these experiments, only 3 epochs in \prsm is sufficient to achieve the best performance in the cryptocurrency application, so further cryptographic optimizations like verification sharding~\cite{CNG18,SDM19} would better leverage \prsm.

\section{Related Work}\label{section:rw}

Research on Byzantine fault tolerance (BFT) started in the 80's~\cite{PSL80,LSP82} and 
later improvements~\cite{CL02} led to a long series of leader-based BFT systems. The technique of pipelining originated in network~\cite{PM95} and was later applied to consensus in the context of crash fault tolerance~\cite{Lam98}.

\paragraph{Bypassing the leader bottleneck.}
\sloppy{The idea that the leader-based design can limit the throughput of consensus is not new~\cite{BS10,BMS12,ABQ13,CNG18,SDM19}.
S-Paxos is a variant of Paxos that aims at disseminating client requests to all replicas to offload the leader~\cite{BMS12}. In particular, it increases the throughput of Paxos by balancing the CPU workload over all replicas, similar to what our hashing phase does in \prsm. However, it 
tolerates only crash failures.
RBFT~\cite{ABQ13} uses multiple concurrent instances of PBFT to detect a slow master instance and triggers a leader replacement through PBFT's complex view change. Mir-BFT~\cite{SDM19} combines these instances to outperform PBFT but when some leaders fail, the throughput can only recover after multiple view changes discard the faulty leaders.
Some consensus algorithms alleviate the need for a leader by requiring an exponential information gathering tree~\cite{BS10} or synchrony~\cite{GHM17}. 
Democratic BFT does not use a leader but a weak coordinator whose failure does not prevent termination within the same round~\cite{CNG18}.
Despite this observation, most blockchain \smr{}s build upon the classic leader-based pattern~\cite{Buc16,EGSvR16,KJGK16,KJGG17,GAGM18}.}

EPaxos~\cite{MAK13} bypasses the leader whose failure might impact performance by exploiting 
one leader per command issued. If these commands do not conflict, they are committed concurrently.
Atlas~\cite{EBR20} improves over EPaxos on 13 sites distributed world-wide by adding replicas closer to clients. These solutions are not Byzantine fault tolerant.
Classic reductions~\cite{BCG93} from the multivalue consensus to the binary consensus problem, like the one we use, avoid the leader to propose the value that will be decided. Instead they reliably broadcast values and spawn binary consensus instances, which has already proved efficient in \smr{}s~\cite{MSC16,CNG18}.
HoneyBadger~\cite{MSC16} is an \smr for asynchronous networks building upon this reduction. It exploits erasure coding to limit the communication complexity of the reliable broadcast.
As consensus cannot be solved in an asynchronous model~\cite{FLP85}, it builds upon a randomized consensus~\cite{MMR14} that converges in constant expected time as long as messages are scheduled in a fair manner~\cite{MMR14,TG19}, an assumption called fair scheduler~\cite{BT85}.
%
Red Belly Blockchain~\cite{CNG18} is deterministic, works in a partially synchronous model~\cite{DLS88}  and outperforms HoneyBadger by avoiding the CPU overhead of erasure coding 
and introducing verification sharding.
Although it relies on DBFT~\cite{CGLR18} like we do to balance the load across
multiple links, it does not leverage the bandwidth like \prsm because it does not pipeline: consensus instances are executed sequentially as each block depends on the previous one.
%


\ifnottechrep\vspace{-0.8em}\fi
\paragraph{Pipelining to increase performance.}
The idea of pipelining is quite old and consists, in the context of networking, of sending a packet before its predecessor has been acknowledged within the same connection~\cite{PM95}.
The original version of Paxos~\cite{Lam98} mentioned the idea of pipelining as ballotting could take place in parallel with ballots initiated by different legislators. Pipelining ballots in Paxos has also been implemented~\cite{KSZ11,SS12,SS13}, however, the benefit of pipelining was not significant.
JPaxos~\cite{KSZ11} is an implementation of MultiPaxos that focuses on recovery, batching, pipelining and concurrency, but the pipeline is actually tuned in a separate work~\cite{SS12,SS13} where the leader of MultiPaxos can spawn multiple instances in parallel to increase the resource utilization. The authors note that the drawback of the approach is that pipelining may lead to congestion: multiple instances can max out the leader's CPU or cause network congestion. 
Distributed pipelining leads to different conclusions.

Chain~\cite{AGK14} organizes nodes in a different pipeline so that only one head node, that can be seen as a leader, spawns all instances.
\hotstuff~\cite{YMR19} is a leader-based \smr with a reduced communication complexity. Its leader piggybacks the phase of one consensus instance with the phase of another consensus instance, hence offering a form of pipelining. In addition it reduces the leader bottleneck by having clients sending their proposals directly to all replicas so that the leader can simply send digests in all the consensus phases. To reduce message complexity, \hotstuff makes use of threshold signatures.
\prsm differs in that it does not reduce message complexity but instead balances its network load by having any replica spawn new consensus instances based on its resource usage.

The COP scheme~\cite{BDK15} dispatches batches to concurrent consensus instances to execute pipeline stages in parallel at every replica by exploiting the multiple cores available.  \prsm distributed the pipeline over the network, minimizing the number of threads per replica in an even-driven loop that favors run-to-completion. Multiple threads are only spawned for cryptographic tasks including the hashing task.

Mencius~\cite{MJM08} is an \smr that exploits pipelining by allowing replicas that are faster than the leader to propose a \code{no-op} proposals, hence allowing to speed up the consensus by (i)~preventing replicas with nothing to propose from blocking the protocol and (ii)~coping with faulty leaders. The authors mentioned the difficulty of making Mencius Byzantine fault tolerant because not all communications are exchanged through a quorum and would require a trusted component. 
%
Some efforts were devoted to reducing the latency of replicated state machines to increase their throughput without the need for pipelining~\cite{FV97,SB12,MBS13}. 
\bftmencius~\cite{MBS13} consists of upper-bounding the latency of updates initiated by correct processes by using an abortable timely announced broadcast. 
Like in Mencius, BFT-Mencius allows a replica to skip its turn by proposing \code{no-op}. 
The experimental results clearly indicates in a cluster setting that the lower the latency the higher the throughput for BFT-Mencius as well as Spinning and PBFT~\cite{CL02}. This confirms previous observations on non-pipelined replicated state machines~\cite{FV97}. Some research work even explored the latency optimality of BFT state machine replication~\cite{SB12}, which was implemened in \bftsmart~\cite{BSA14}.
Our approach indicates that distributed pipelining can lead to significantly higher throughput while still achieving reasonable latency at large scale.

\ifnottechrep\vspace{-1.2em}\fi
\section{Conclusion}\label{section:conclusion}
\prsm is the first \smr to exploit resources through distributed pipelining.
At 128 nodes, it improves the throughput of the HotStuff \smr we know by 12-fold. 
Within a cryptocurrency application, \prsm demonstrates that blockchains can suffer from bottlenecks that are not related to the network usage of the consensus protocols they build upon.
Our work has revealed the significant impact of the HOL blocking factor on performance, that was neglected because it could not be observed in slower \smr{}s before. This observation opens up new interesting research directions on the choice of layer-4 network protocols for large-scale applications like blockchains.

\iftechrep
\section*{Acknowledgments}
\sloppy{We wish to thank Alysson Bessani for his fruitful comments on an earlier version of this draft. This research is supported under Australian Research Council Discovery Projects funding scheme (project number 180104030) entitled ``Taipan: A Blockchain with Democratic Consensus and Validated Contracts'' and Australian Research Council Future Fellowship funding scheme (project number 180100496) entitled ``The Red Belly Blockchain: A Scalable Blockchain for Internet of Things''.}
\fi

\bibliographystyle{plain}
\bibliography{references}


\end{document}